\documentclass[12pt]{iopart}

\newif\ifanonymized
\anonymizedfalse

\newif\ifincludesupplement
\includesupplementtrue

\newif\ifshowupdates
\showupdatesfalse

\expandafter\let\csname equation*\endcsname\relax
\expandafter\let\csname endequation*\endcsname\relax

\usepackage{amsmath}
\usepackage{amssymb}
\usepackage{csquotes}
\usepackage{graphicx}
\usepackage{dcolumn}
\usepackage{bm}
\usepackage{hyperref}
\usepackage{xcolor}
\usepackage{tabularx}
\usepackage{booktabs}
\usepackage{cite}
\usepackage{tikz}
\usepackage{fancyhdr}

\newcommand{\avgx}[2]{\left\langle{#1}\right\rangle_{#2}}
\newcommand{\abs}[1]{\left|{#1}\right|}
\newcommand{\kl}[2]{D\!\left({#1}\|{#2}\right)}
\newcommand{\nlog}[1]{\ln{#1}}
\newcommand{\set}[1]{\{#1\}}
\newcommand{\brk}[1]{\!\left(#1\right)}
\ifshowupdates
\newcommand{\update}[1]{\textcolor{red}{#1}}
\else
\newcommand{\update}[1]{#1}
\fi

\begin{document}

\setcounter{footnote}{6}

\title{Normalizing flows for atomic solids}
\ifanonymized
\author{Anonymous authors}
\else
\author{Peter Wirnsberger\footnotemark{}, George Papamakarios\footnotemark[\value{footnote}], Borja Ibarz\footnotemark[\value{footnote}], S{\'e}bastien Racani{\`e}re, Andrew J.~Ballard, Alexander Pritzel, Charles Blundell
\footnotetext{Authors contributed equally.}}
\address{DeepMind, London, United Kingdom}
\fi
\date{\today}

\begin{abstract}
We present a machine-learning approach, based on normalizing flows, for modelling atomic solids. Our model transforms an analytically tractable base distribution into the target solid without requiring ground-truth samples for training. We report Helmholtz free energy estimates for cubic and hexagonal ice modelled as monatomic water as well as for a truncated and shifted Lennard-Jones system, and find them to be in excellent agreement with literature values and with estimates from established baseline methods. We further investigate structural properties and show that the model samples are nearly indistinguishable from the ones obtained with molecular dynamics. Our results thus demonstrate that normalizing flows can provide high-quality samples and free energy estimates \update{without the need for multi-staging}.
\end{abstract}

\section{\label{sec:intro}Introduction}
Accurate estimation of equilibrium properties of a thermodynamic system is a central challenge of computational statistical mechanics~\cite{Tuckerman2019, FrenkelSmit}. For decades, molecular dynamics and hybrid Monte Carlo~\cite{Duane1987} have been the methods of choice for sampling such systems at scale~\cite{Yu2016, Hudait2017, Mosalaganti2021}. Recently there has been a surge in using deep learning to develop learned schemes for sampling from probability distributions in general and physical systems in particular, most notably using normalizing flows~\cite{Tabak2013, Rezende2015}. Flow-based learned sampling schemes have been applied to various physical systems, from lattice field theories~\cite{Albergo2019, Boyda2021, Nicoli2021}, to spin systems~\cite{Nicoli2020}, to proteins~\cite{Noe2019}.

Normalizing flows are appealing because of the following two properties: first, they can generate independent samples efficiently and in parallel; second, they can provide the exact probability density of their generation mechanism~\cite{Papamakarios2021, Kobyzev2021}.
Thus, training a flow-based model $q$ to approximate a target distribution $p$ (for example, the Boltzmann distribution of a physical system) yields an efficient but approximate sampler for $p$; re-weighting the samples by their probability density (for example, using importance sampling) can then be used to remove estimation bias \cite{Bugallo2017, Muller2019, Noe2019}.
For free energy estimation in particular, flows are interesting because they do not require samples from intermediate thermodynamic states to obtain accurate estimates, unlike traditional estimators such as thermodynamic integration~\cite{FrenkelSmit} or the multistate Bennett acceptance ratio (MBAR) method~\cite{Shirts2008}. Instead, the flow model can be used as part of a targeted
estimator~\cite{Jarzynski2002, Hahn2009, Wirnsberger2020, Nicoli2020, Ding2020, Rizzi2021, Xinqiang2021, Nicoli2021} which was demonstrated to be competitive to MBAR in terms of accuracy when applied to a small-scale solvation problem~\cite{Wirnsberger2020}.

Despite their appeal for both sampling and free energy estimation of atomistic systems, constructing and training flow-based models that can rival the accuracy of already established methods remains a significant challenge. One of the reasons is that for simple re-weighting schemes such as importance sampling to be accurate in high dimensions, the model $q$ must be a very close approximation to the target distribution $p$, which is hard to achieve with off-the-shelf methods.
Even for common benchmark problems of identical particles, such as a Lennard-Jones system~\cite{FrenkelSmit}, successful training has thus far been demonstrated for small system sizes of up to tens of particles (for example, 38 particles in Ref.~\cite{Noe2019}, 13 particles in Ref.~\cite{Kohler2020}), requiring ground-truth samples from $p$ to train in all cases. Addressing this limitation is crucial for scaling up flow-based methods to systems of interest in statistical mechanics.

In this work, we propose a flow model that is tailored to sampling from atomic solids of identical particles, and we demonstrate that it can scale to system sizes of up to 512 particles with excellent approximation quality.
The model is trained \update{to approximate the Boltzmann distribution of a chosen metastable solid by fitting against a known potential energy function. Training uses only the energy evaluated at model samples, and does not require samples from the Boltzmann distribution as ground truth}. We examine the quality of the learned sampler using a range of sensitive metrics and estimate Helmholtz free energies of a truncated and shifted Lennard-Jones system (FCC phase) and of ice I (cubic and hexagonal) using a monatomic model~\cite{Molinero2009}. Comparison with baseline methods shows that our flow-based estimates are highly accurate, allowing us to resolve small free energy differences.

\section{\label{sec:method}Method}
We begin by considering a system of $N$ identical particles interacting via \update{a known} energy function $U$ and attached to a heat bath at temperature $T$. The equilibrium distribution of this system is given by the Boltzmann distribution
\begin{equation}
p(x) = \frac{1}{Z}\exp[-\beta U(x)],    
\end{equation}
where $x$ denotes a point in the $3N$-dimensional configuration space, $Z=\int \mathrm dx \exp[-\beta U(x)]$ is the partition function, $\beta = 1/k_\mathrm{B} T$ is the inverse temperature and $k_\mathrm{B}$ is the Boltzmann constant. Our aim is to build and train a flow model that can accurately approximate $p$, for atomic solids in particular.

\subsection{Flow models}

A flow model is a probability distribution $q$ defined as the pushforward of an analytically tractable base distribution $b$ through a flexible diffeomorphism $f$, typically parameterized by neural networks \cite{Papamakarios2021, Kobyzev2021}. Independent samples from $q$ can be generated in parallel, by first sampling $z$ from $b$ and taking $x = f(z)$. The probability density of a sample can be obtained using a change of variables,
\begin{equation}
    q(x) = b(z)\abs{\det J_f(z)}^{-1},
\end{equation}
where $J_f$ is the Jacobian of $f$. 

We can train $q$ to approximate $p$ by minimizing a loss function that quantifies the discrepancy between $q$ and $p$. In this work, we use the following Kullback--Leibler divergence as the loss function
\begin{eqnarray}
    \label{eq:kl}
    \kl{q}{p} &= \avgx{\nlog{q(x)} - \nlog{p(x)}}{q} \\ 
    &= \avgx{\nlog{q(x)} + \beta U(x)}{q} + \nlog{Z}  \nonumber \\
    &= \avgx{\nlog{b(z)} - \nlog{\abs{\det J_f(z)}} + \beta U(f(z))}{b} + \nlog{Z} \nonumber.
\end{eqnarray}
Since $\nlog{Z}$ is a constant with respect to the parameters of $q$, it can be ignored during optimization. The expectation in the final expression can be estimated using samples from $b$, so $\kl{q}{p}$ can be minimized using stochastic gradient-based methods. This loss function is appealing because it does not require samples from $p$ or knowledge of $Z$, and can be optimized solely using evaluations of the energy $U$.

\subsection{Systems and potentials}

The systems considered in this work are crystalline solids consisting of $N$ indistinguishable atoms at constant volume and temperature. The crystal is assumed to be contained in a $3$-dimensional box with edge lengths $L_1, L_2, L_3$ and periodic boundary conditions. A configuration $x$ is an $N$-tuple $(x_1, \ldots, x_N)$, where $x_n = (x_{n1}, x_{n2}, x_{n3})$ are the coordinates of the $n$-th atom and $x_{ni} \in [0, L_i/\sigma]$ with $\sigma$ being a characteristic length scale of the system (here the particle diameter). By expressing $x$ in reduced units, $Z$ and all probability densities become dimensionless.

The potentials used in this work are invariant to global translations (with respect to periodic boundary conditions) and arbitrary atom permutations. Both of these symmetries
are incorporated into the model architecture; that is, we design the base distribution $b$ and the diffeomorphism $f$ such that the density function $q$ is invariant to translations and atom permutations, as well as compatible with periodic boundary conditions. We explain how this is achieved in the following paragraphs.

\subsection{Model architecture}

\begin{figure}
\centering
\includegraphics[width=0.7\textwidth]{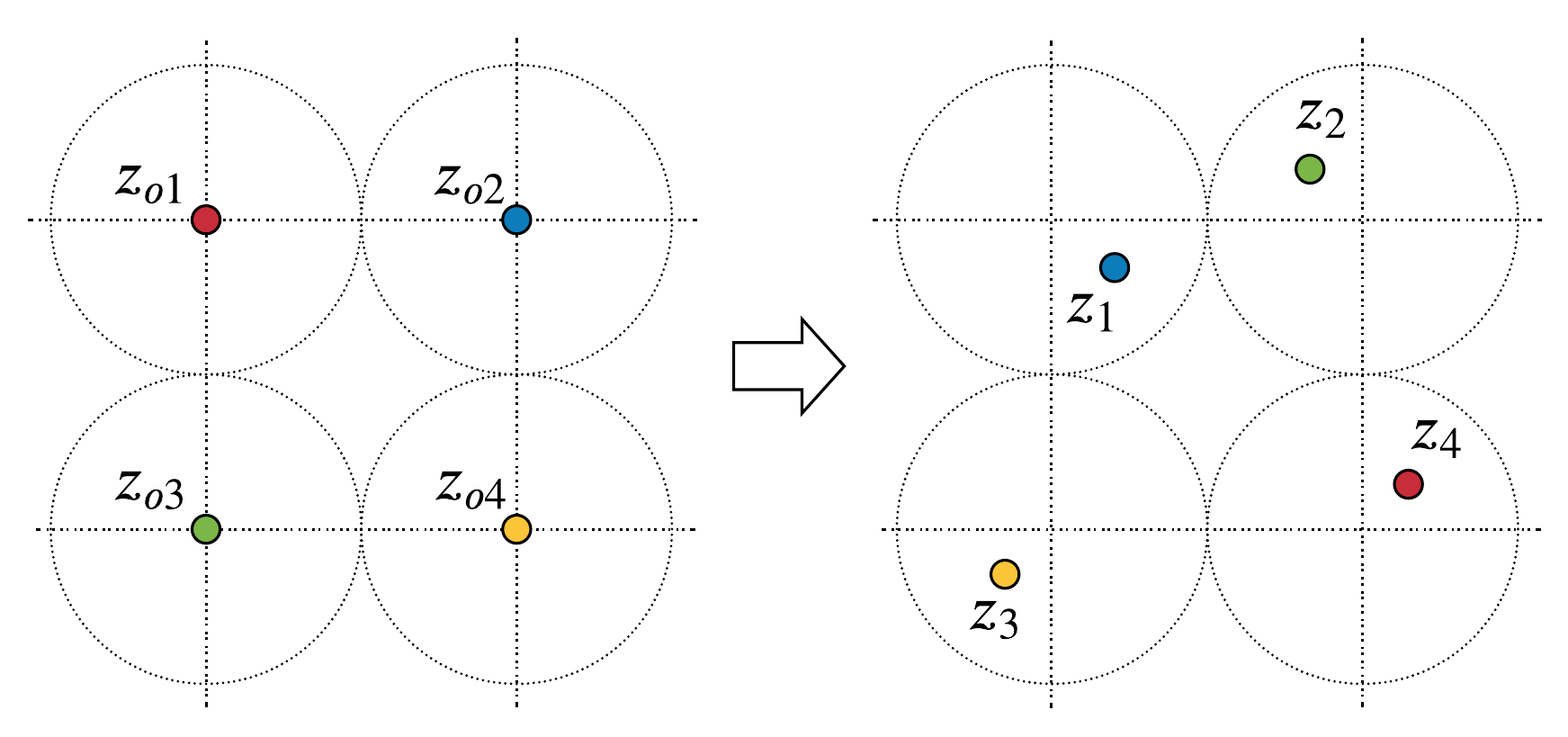}
\caption{\label{fig:base_lattice} Illustration of generating samples from the base distribution in 2D: atoms arranged on a lattice (left) are perturbed with truncated Gaussian noise and randomly permuted (right).} 
\end{figure}

Our base distribution $b$ is constructed as a lattice with $N$ sites that have been randomly perturbed and permuted as illustrated in Fig.~\ref{fig:base_lattice}. Starting from a lattice $z_o = (z_{o1}, \ldots, z_{oN})$, a configuration $z = (z_{1}, \ldots, z_{N})$ is generated by independently adding spherically-truncated Gaussian noise to each lattice site, followed by a random permutation of all atoms. The truncation is chosen such that no two neighbouring atoms can swap lattice sites, so that after the permutation, all atoms can be traced back to the site they originated from. This construction yields a base distribution that can be trivially sampled from, and has a permutation-invariant probability density function that can be evaluated exactly.

Our diffeomorphism $f$ is implemented as a sequence of $K$ invertible functions composed such that $f = f_K \circ \cdots \circ f_1$. Each function $f_k$ is parameterized by a separate neural network, whose parameters are optimized by the loss in Eq.~\eqref{eq:kl}. We implement the functions $f_k$ using an improved version of the model proposed in Ref.~\cite{Wirnsberger2020}. In this model, each $f_k$ transforms element-wise either one or two coordinates of all atoms as a function of all remaining coordinates. The transformation of each coordinate is implemented using circular rational-quadratic splines~\cite{Rezende2020}, which ensures that the transformation is nonlinear, invertible and obeys periodic boundary conditions. The spline parameters are computed as a function of the remaining coordinates using multiple layers of self-attention, a neural-network module commonly used in language modelling \cite{Vaswani2017}. Ref.~\cite{Wirnsberger2020} showed that a diffeomorphism $f$ parameterized this way is equivariant to atom permutations, which means that permuting the input of $f$ has the same effect as permuting the output of $f$. As also shown by Refs.~\cite{Bender2020, Kohler2020}, the combination of a permutation-invariant base distribution with a permutation-equivariant diffeomorphism yields a permutation-invariant distribution $q$, as desired. More implementation details are provided in the Supplementary Material.

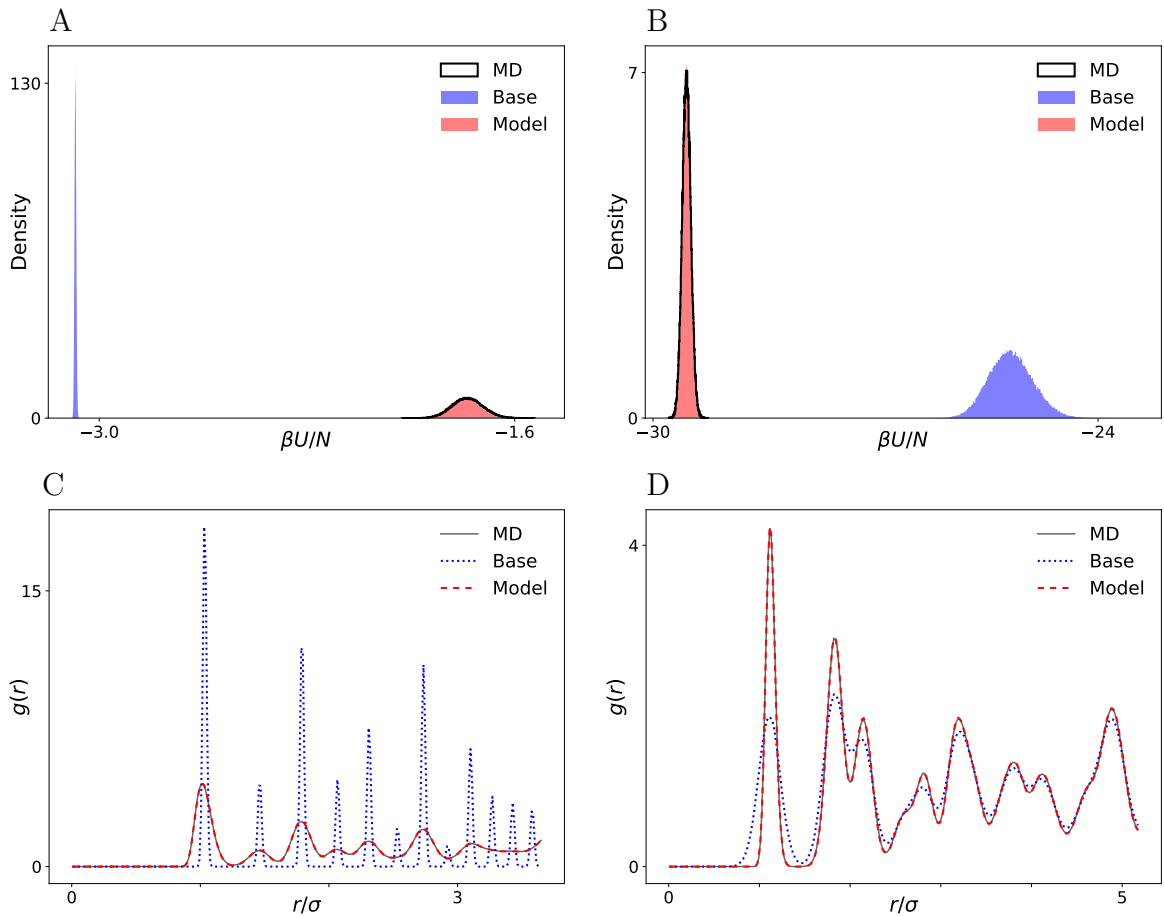
\begin{figure*}
    \begin{tikzpicture} [every node/.style={inner sep=0,outer sep=-1}]
        \draw (0, 0) node[inner sep=0] {\includegraphics[width=.5\textwidth]{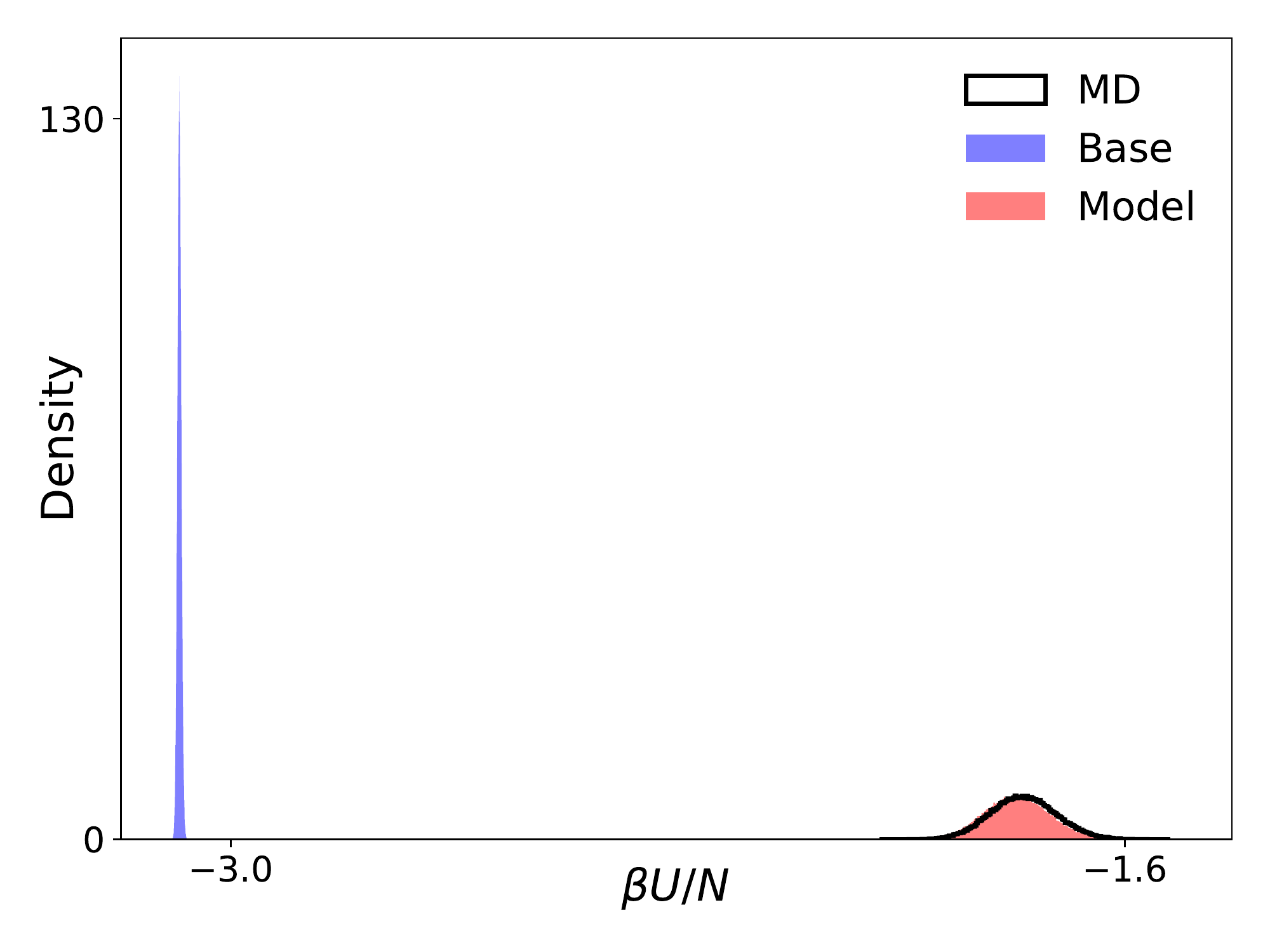}};
        \coordinate (A) at (current bounding box.north west);
        \node[above right, xshift=0.8cm] at (A) {A};
    \end{tikzpicture}
    \begin{tikzpicture} [every node/.style={inner sep=0,outer sep=-1}]
        \draw (0, 0) node[inner sep=0] {\includegraphics[width=.5\textwidth]{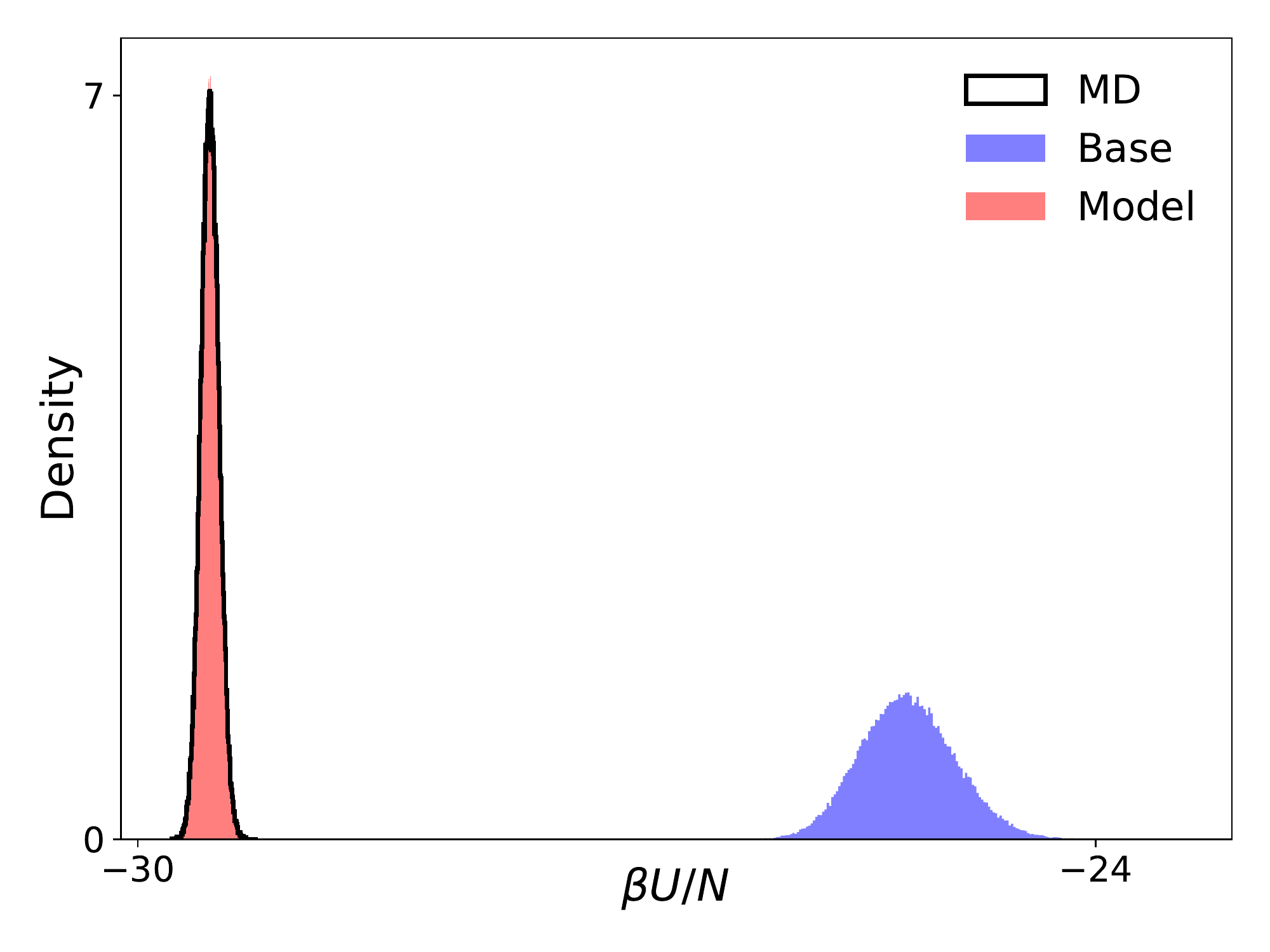}};
        \coordinate (B) at (current bounding box.north west);
        \node[above right, xshift=0.8cm] at (B) {B};
    \end{tikzpicture}
    \begin{tikzpicture} [every node/.style={inner sep=0,outer sep=-1}]
        \draw (0, 0) node[inner sep=0] {\includegraphics[width=.5\textwidth]{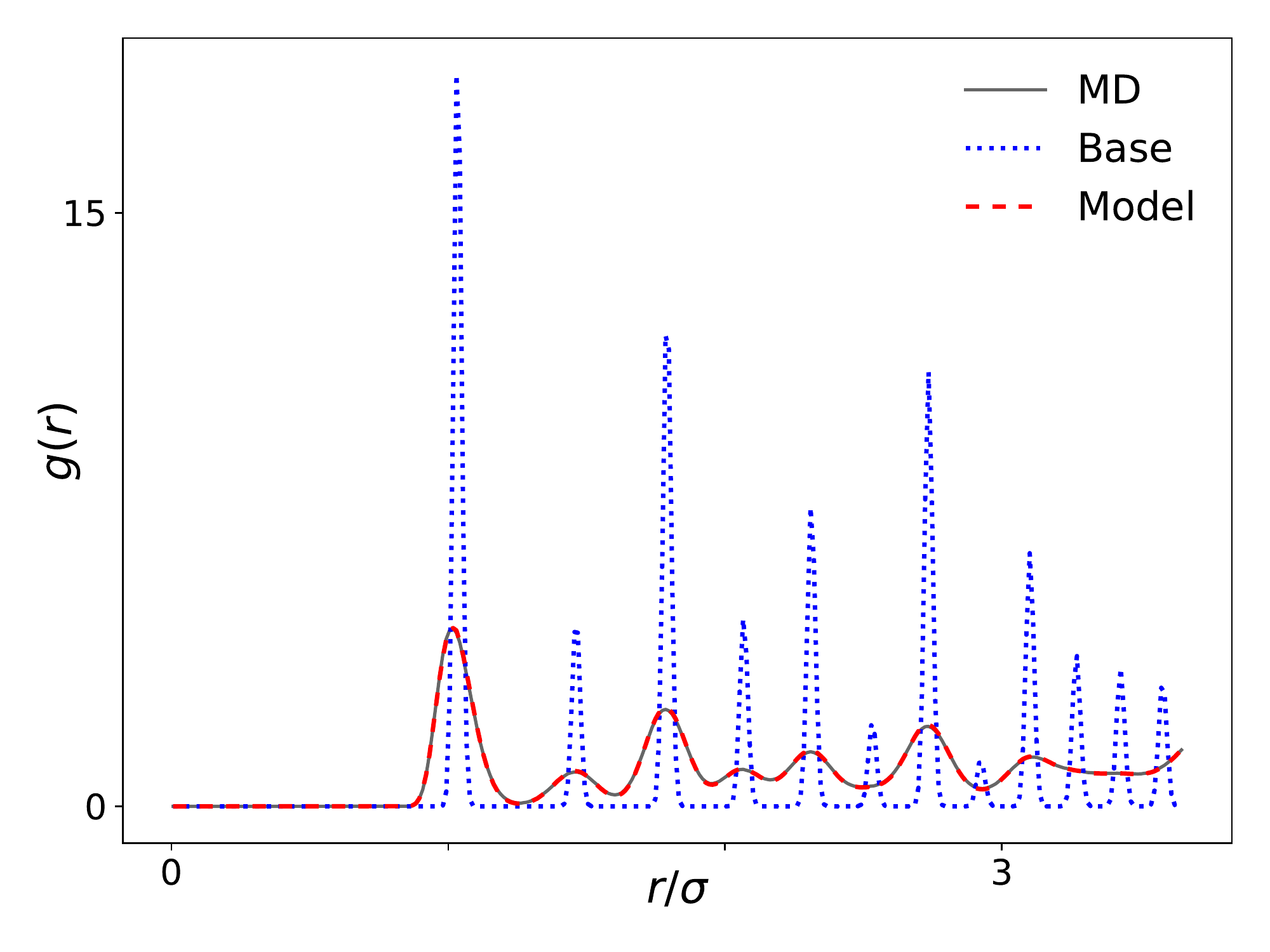}};
        \coordinate (C) at (current bounding box.north west);
        \node[above right, xshift=0.7cm] at (C) {C};
    \end{tikzpicture}
    \begin{tikzpicture} [every node/.style={inner sep=0,outer sep=-1}]
        \draw (0, 0) node[inner sep=0] {\includegraphics[width=.5\textwidth]{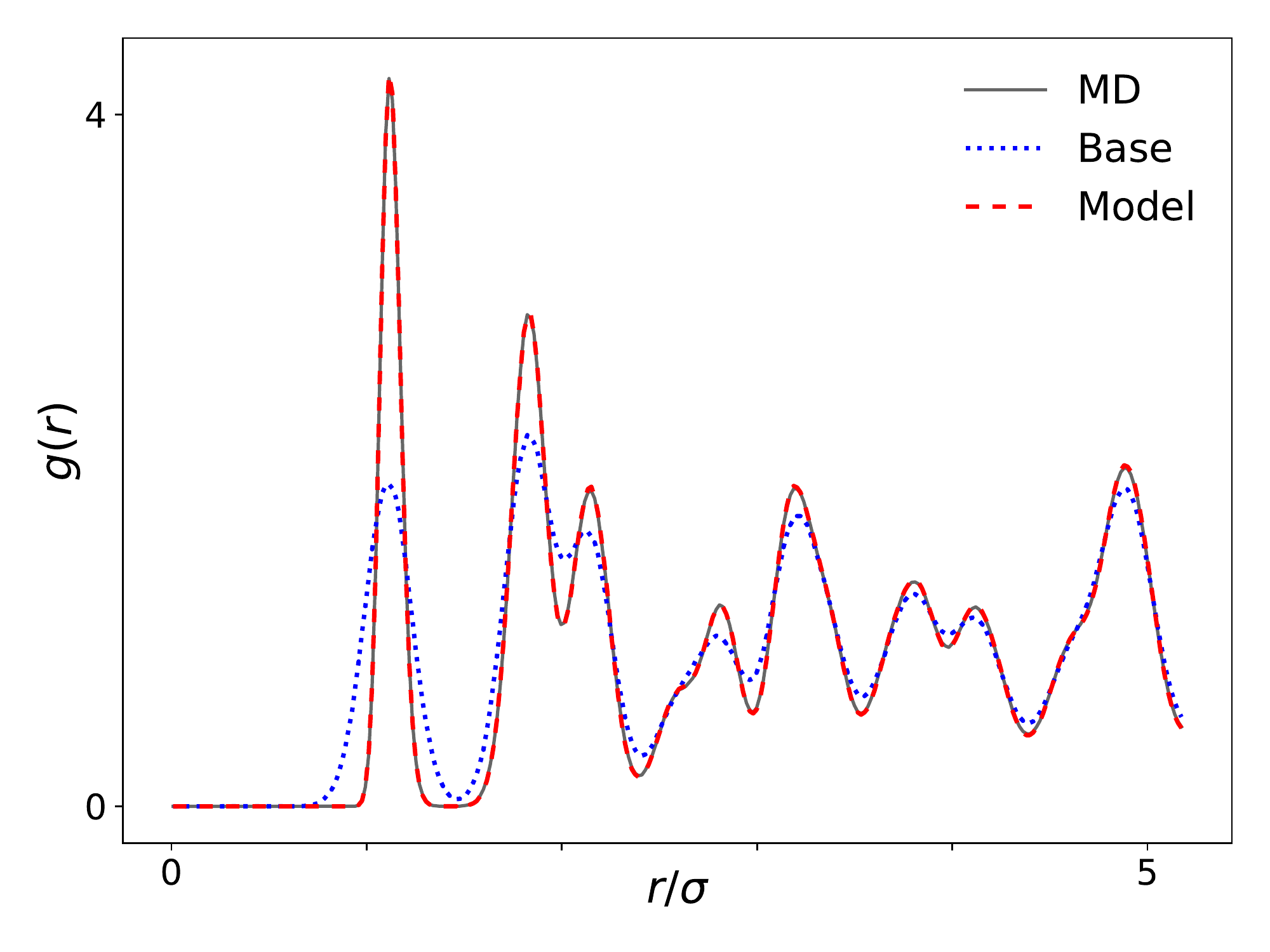}};
        \coordinate (D) at (current bounding box.north west);
        \node[above right, xshift=0.8cm] at (D) {D};
    \end{tikzpicture}
\caption{\label{fig:rdf}Energy histograms (A--B) and radial distribution functions (C--D) of the base distribution, the fully trained model and MD simulation data, for the 500-particle LJ system (left) and the 512-particle cubic ice system (right).}
\end{figure*}

Finally, we incorporate the translational symmetry into the above architecture as follows. We fix the coordinates of an arbitrary reference atom (say $x_1$), and use the flow model to generate the remaining $N-1$ atom coordinates as described above. Then, we globally translate the atoms uniformly at random (under periodic boundary conditions), so that the reference atom can end up anywhere in the box with equal probability. Since the index of the reference atom and its original position are known and fixed, we can reverse this operation in order to obtain the probability density of an arbitrary configuration. This procedure yields a translation invariant probability density function that can be calculated exactly.

A key feature of our model architecture is that we can target specific crystal structures by encoding them into the base distribution. For example, if we are interested in modelling the hexagonal phase of a crystal, we can choose the lattice of the base distribution to be hexagonal. Empirically, we find that, after training, the flow model becomes a sampler for the (metastable) crystal state that we encode in the base distribution, and does not sample configurations from other states. Thus, by choosing the base lattice accordingly, we can guide the model towards the state of interest, without changing the energy function or using ground-truth samples for guidance.

\section{\label{sec:results}Results}
We train the models on two different systems. The first is a truncated and shifted Lennard-Jones (LJ) crystal in the FCC phase at reduced temperature and density values of 2 and 1.28, respectively, employing a reduced cutoff of 2.7 as in Ref.~\cite{Aragones2012}. The second is ice I modelled as monatomic Water (mW)~\cite{Molinero2009} in the diamond cubic (Ic) and hexagonal (Ih) phases at a temperature of 200~K and a density of approximately $1.004~\mathrm{g}/\mathrm{cm}^3$ similar to Ref.~\cite{Quigley2014}. All further simulation details are provided in the Supplementary Material.

\update{Code for reproducing the experiments and pre-trained models are provided at
\ifanonymized
\url{https://github.com/<anonymous-account>/flows_for_atomic_solids}.
\else
\url{https://github.com/deepmind/flows_for_atomic_solids}.
\fi
The code uses JAX \cite{jax2018github}, Haiku \cite{haiku2020github} and Distrax \cite{deepmind2020jax} for building and training models.}

\subsection{Evaluation of model samples}

To assess the quality of our trained models, we compare against molecular dynamics (MD\@). We ran \textit{NVT} simulations of the target systems using the simulation package LAMMPS~\cite{Thompson2022}.
Figure~\ref{fig:rdf} shows the energy histogram and radial distribution function (RDF) of the Lennard-Jones FCC crystal and of cubic ice, as computed by samples from the base distribution, the trained model, and MD\@. \update{The RDF $g(r)$ is the ratio of the average number density at a distance $r$ of an arbitrary reference atom and the average number density in an ideal gas at the same overall density~\cite{FrenkelSmit}.} By construction, the base distribution captures the locations of the peaks in the RDF correctly, but its energy histogram is far off compared to the MD result. The energy histograms and the RDFs computed from model samples, however, are nearly indistinguishable from MD in both systems\@.
Importantly, no unbiasing or re-weighing was necessary to obtain this quality of agreement. The results for hexagonal ice show a similar level of agreement (see Supplementary Material). These results demonstrate that the mapping $f$ successfully transforms the base distribution into an accurate sampler.

\begin{figure}
\centering
\includegraphics[width=0.6\textwidth]{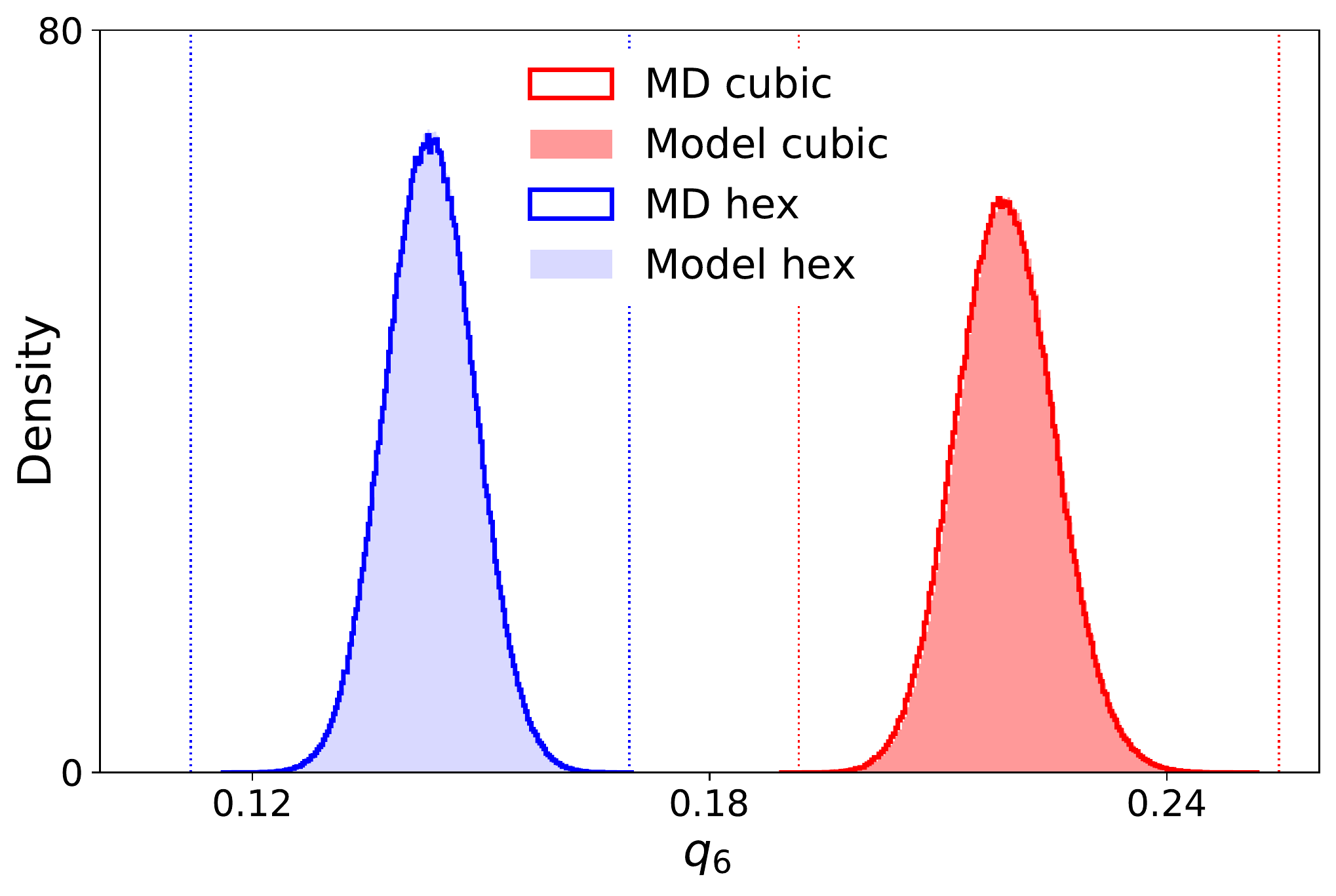}
\caption{\label{fig:bop} Histograms of averaged local bond order parameter $q_6$ for 512-particle cubic and hexagonal ice. Solid lines indicate the histograms from MD samples; shaded areas from model samples. The vertical lines mark the maximum and minimum values seen in 1M bond order parameters for each of the two model histograms.}
\end{figure}

To demonstrate that the trained model becomes a sampler of the (metastable) crystal structure that we have encoded into the base distribution, we computed histograms of the averaged local bond-order parameter $q_6$~\cite{Lechner2008} that was designed to discriminate between different phases (see Fig.~\ref{fig:bop}). We can see that the two histograms are well separated and agree with MD, showing that the model does indeed become an accurate sampler of only the crystal state encoded in the base distribution. If that were not the case, we could have enforced it by adding a biasing potential to the energy, as commonly done in nucleation studies that employ umbrella sampling~\cite{Auer2001}.

\subsection{Free energy estimation}

Next, we use the trained flow models to estimate the Helmholtz free energy $F$ for various system sizes, which is given by~\cite{FrenkelSmit}
\begin{equation}
F = -\beta^{-1}\left(\nlog{Z}-\nlog{N!}\right).
\end{equation}
We note that the thermal de Broglie wavelength does not appear in the above expression 
as we set it to $\sigma$, following Ref.~\cite{Vega2007}, and absorb it into $Z$ by expressing $x$ in reduced units. We then estimate $\nlog{Z}$ by first defining a generalized work function~\cite{Jarzynski2002}
\begin{equation}
  \label{eq:phi}
  \beta \Phi(x) = \beta U(x) + \nlog{q(x)}.
\end{equation}
The average work value $\avgx{\beta \Phi(x)}{q}$ is also our training objective in Eq.~\eqref{eq:kl}. We then harness the trained flow to draw a set of samples $\set{x^{(m)}}_{m=1}^M$ from $q$ and estimate $\nlog{Z}$ via the targeted free energy perturbation estimator~\cite{Jarzynski2002}
\begin{equation}
\label{eq:lfep}
\nlog{Z} = \nlog{\avgx{\exp(-\beta \Phi(x))}{q}} \approx \nlog{\frac{1}{M}\sum_{m=1}^M\exp\brk{-\beta\Phi(x^{(m)})}}.
\end{equation}
The above estimation method, referred to as learned free energy perturbation (LFEP) in combination with a learned model~\cite{Wirnsberger2020}, is appealing, because it does not require samples from $p$ for either training the model or for evaluating the estimator.

\begin{figure}
\centering
    \begin{tikzpicture} [every node/.style={inner sep=0,outer sep=-1}]
        \draw (0, 0) node[inner sep=0] {\includegraphics[width=.495\textwidth]{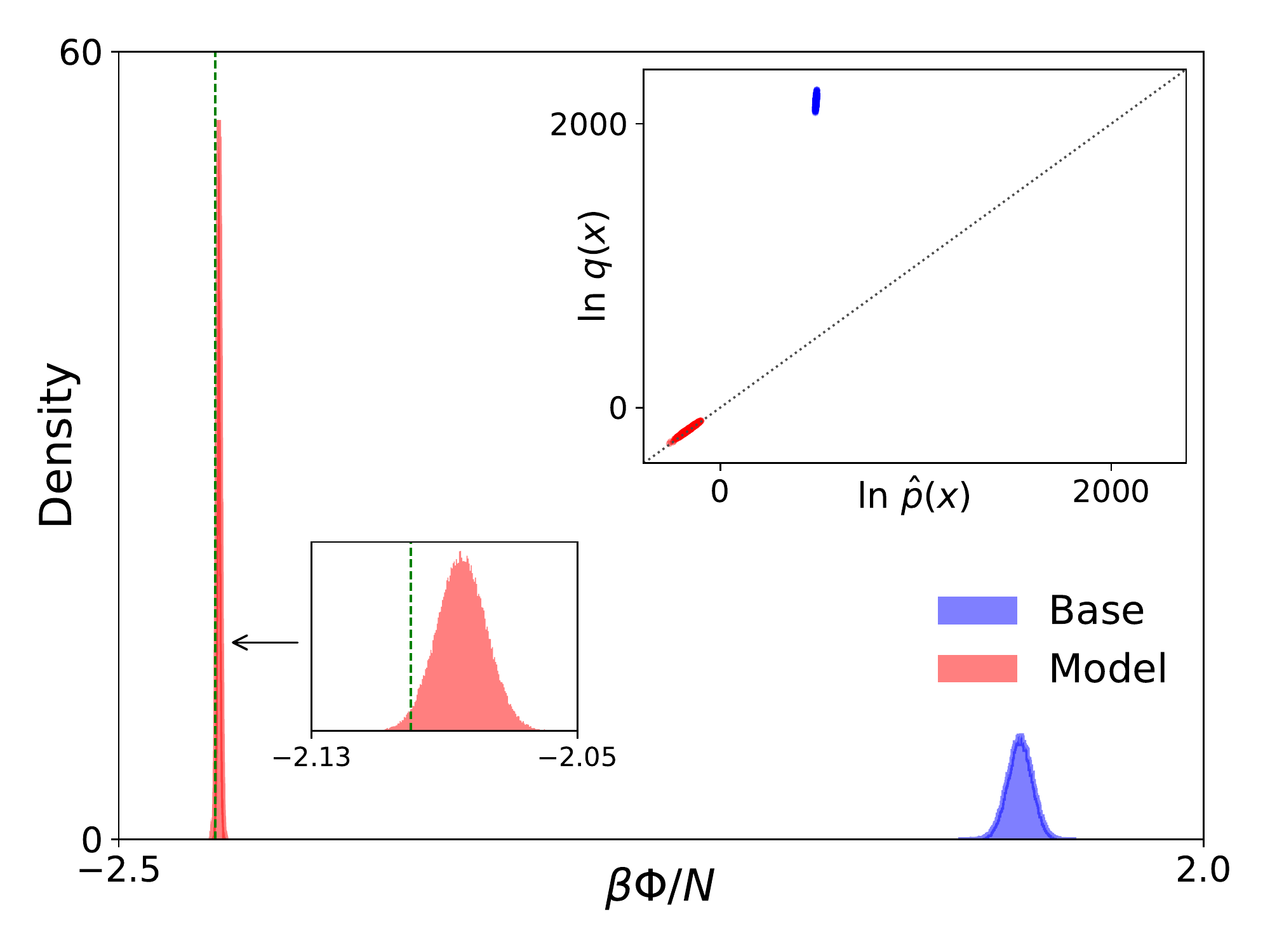}};
        \coordinate (A) at (current bounding box.north west);
        \node[above right, xshift=0.8cm] at (A) {A};
    \end{tikzpicture}
    \begin{tikzpicture} [every node/.style={inner sep=0,outer sep=-1}]
        \draw (0, 0) node[inner sep=0] {\includegraphics[width=.495\textwidth]{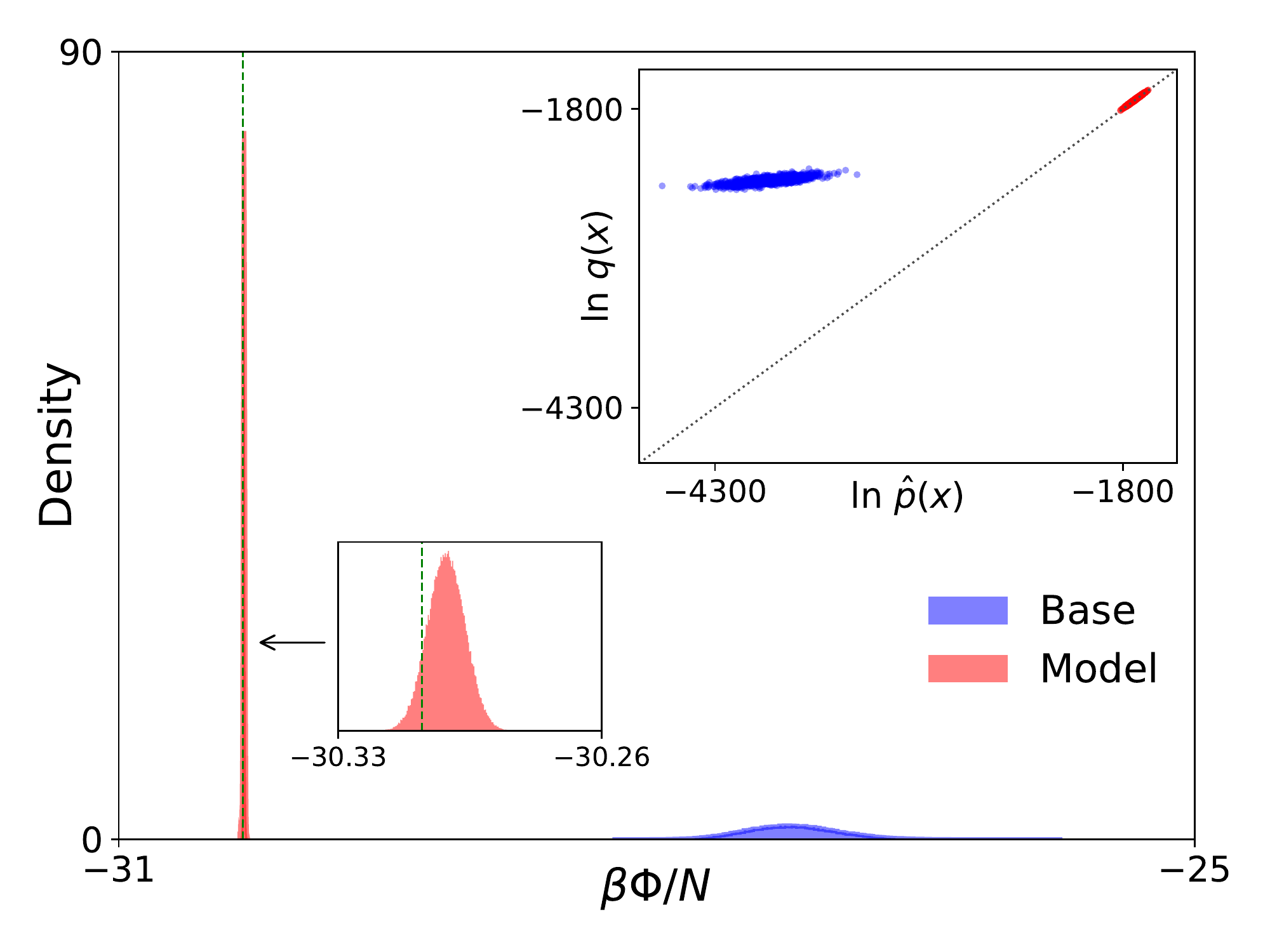}};
        \coordinate (B) at (current bounding box.north west);
        \node[above right, xshift=0.8cm] at (B) {B};
    \end{tikzpicture}
\caption{\label{fig:energy} Histograms of work values ($\beta\Phi$) per particle from base and model samples for 500-particle LJ crystal (A) and 512-particle cubic ice (B). The dashed vertical line marks $-\widehat{\nlog{Z}}/N$ as estimated by MBAR (enlarged in inset). Upper insets: scatter plots of model density vs approximately normalized target density $\ln \hat{p} = -\beta U - \widehat{\nlog{Z}}$ computed from model samples, where the model is either the base distribution or the fully trained model and $\widehat{\nlog{Z}}$ is the MBAR estimate. \update{The dotted diagonal marks the identity.}}
\end{figure}
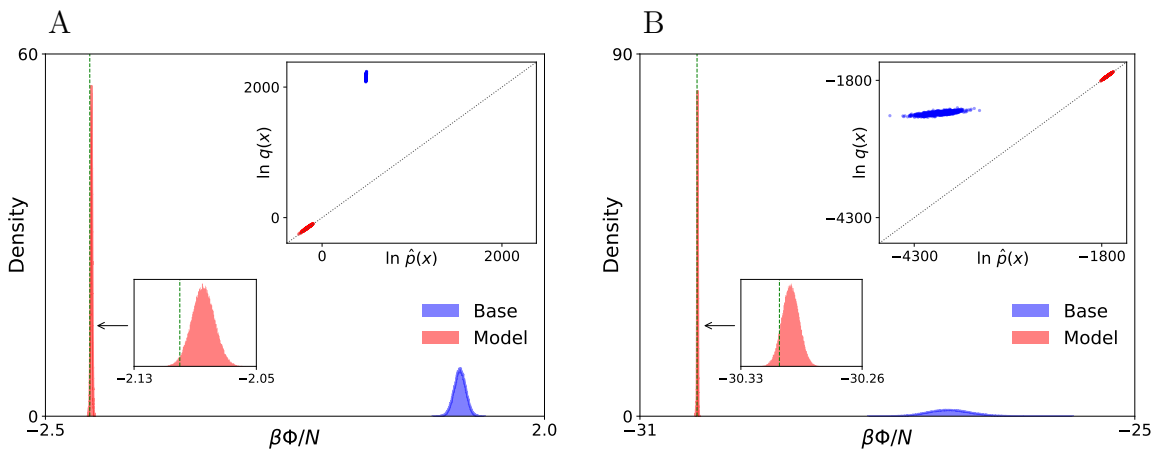

Although the approximation in Eq.~\eqref{eq:lfep} becomes exact in the limit of an infinite sample size, to obtain accurate results for a finite sample set we require sufficient agreement between the proposal and the target distributions~\cite{Jarzynski2002, Wirnsberger2020}. Since $\beta \Phi$ quantifies the pointwise difference between $\nlog{q}$ and $\nlog{p}$ (up to an additive constant), the distribution of generalized work values is a good metric for assessing the quality of the flow for free energy estimation.

Figure~\ref{fig:energy} compares distributions of work values computed for the base distribution and for the fully trained model.  \update{From the non-negativity of the Kullback--Leibler divergence in Eq.~\eqref{eq:kl} it follows that $\avgx{\beta \Phi(x)}{q} \ge -\nlog{Z}$, with equality if and only if $q$ and $p$ are equal. Therefore, the gap between the average work value and $-\nlog{Z}$ (which we aim to estimate) quantifies how accurately the model $q$ approximates the Boltzmann distribution $p$; for a perfect model, we would expect to see a delta distribution located at $-\nlog{Z}$~\cite{Jarzynski2002}.} The work values obtained with the trained model are indeed sharply peaked near our MBAR estimate of $-\widehat{\nlog{Z}}$. On a qualitative level, this shows a clear benefit of LFEP~\cite{Wirnsberger2020} over the original FEP estimator corresponding to $f$ being the identity map, which failed to converge on this problem.

Using the same trained model as in LFEP, we can also employ a learned version of the 
bidirectional BAR estimator (LBAR)~\cite{Bennett1976, Wirnsberger2020}. LBAR uses samples from both the base distribution $b$ and from the target $p$ and is known to be the minimum variance estimator for any asymptotically unbiased method~\cite{Shirts2003}. The downside compared to LFEP is that an additional MD simulation needs to be performed to obtain samples from $p$ for computing the estimator (but not for training the model).

To verify the correctness of the flow-based free energy estimates, we require an accurate baseline method.
The Einstein crystal (and molecule) method~\cite{Frenkel1984, Vega2007} and lattice-switch Monte Carlo (LSMC)~\cite{Bruce1997} are common choices for computing solid free energies with different trade-offs. While the former is conceptually simple, we found it challenging to optimize it to high precision, which is consistent with previously reported results~\cite{Aragones2012, Vega2007}.
The latter is known to be accurate but provides only free energy differences between two compatible lattices rather than absolute free energies. We therefore tested MBAR as an alternative estimator on this problem and found it yields sufficiently accurate absolute free energy estimates to serve as a reference.

\begin{table}[t]
\caption{\label{tab:fe} 
Helmholtz free energy estimates, $\beta \hat{F}/N$, obtained with 2M, $2\times1$M and $100\times10$k samples for LFEP, LBAR and MBAR for LJ, and 2M, $2\times1$M and $200\times10$k samples for ice. Parentheses show the uncertainties in the last digits (two standard errors); error bars for LFEP and LBAR were computed using $10$ independently trained models, so they quantify uncertainty both due to randomness in training and due to finite sample size in estimation; error bars for MBAR were computed across 10 independent estimates. The literature value for LJ is 3.11(4) for 256 particles~\cite{Aragones2012}\update{; the literature value for mW is unknown.} See Supplementary Material for further details.}
\begin{indented}
\item[]\begin{tabular}{lccll}
\br
System  & $N$      &  LFEP             &  LBAR        &  MBAR           \\ 
\mr
LJ      & 256      & 3.10800(28)       & 3.10797(1)   &  3.10798(9)    \\
LJ      & 500      & 3.12300(41)        & 3.12264(2)   &  3.12262(10)    \\ 
\mr
Ice Ic  & 64       & -25.16311(3)      & -25.16312(1) &  -25.16306(20)   \\
Ice Ic  & 216      & -25.08234(7)      & -25.08238(1) & -25.08234(5)    \\
Ice Ic  & 512      & -25.06163(35)     & -25.06161(1) & -25.06156(3)    \\
\mr
Ice Ih  & 64       & -25.18671(3)     & -25.18672(2) & -25.18687(26)   \\
Ice Ih  & 216      & -25.08980(3)     & -25.08979(1) & -25.08975(14)    \\
Ice Ih  & 512      & -25.06478(9)     & -25.06479(1) & -25.06480(4)    \\
\br
\end{tabular}
\end{indented}
\end{table}

A quantitative comparison of free energy estimates for both systems and different system sizes is shown in Tab.~\ref{tab:fe} (see also Supplementary Material). For LFEP we use $S$ samples from the base distribution $b$; for LBAR we use $S$ samples from $b$, plus another $S$ samples from the target $p$ obtained by MD; for MBAR we use a sufficiently large number of intermediate states between $b$ and $p$ to obtain good accuracy, which we sample using MD\@ (see caption of Tab.~\ref{tab:fe} for exact numbers). Overall, we find excellent agreement of the learned estimators with MBAR for both LJ and ice. The LBAR estimates exhibit lower statistical uncertainties than LFEP across the board, with error bars on the order of $10^{-5} k_\mathrm{B} T$ per particle. We find it remarkable, however, that LFEP can yield comparable accuracy in most cases without access to MD samples for training or estimation, and without the need for defining intermediate states. Finally, \update{we compute} the Helmholtz free energy difference between cubic and hexagonal ice \update{for 216 particles by subtracting the two LFEP estimates in Tab.~\ref{tab:fe}. This yields a value of} $12.4(2)~\mathrm{J/mol}$ \update{which} is in good agreement with the reported Gibbs free energy difference of $11.2(2)~\mathrm{J/mol}$ obtained with LSMC simulations at atmospheric pressure~\cite{Quigley2014}.

\section{\label{sec:discussion}Discussion}
In summary, we have proposed a normalizing-flow model for solids consisting of identical particles and have demonstrated that it can be optimized to approximate Boltzmann distributions accurately for system sizes of up to 512 particles, without requiring samples from the target for training. We have shown that flow-based estimates of radial distribution functions, bond-order parameters and energy histograms agree well with MD results, without the need for an unbiasing step. A detailed comparison of free energy estimates further verifies that our flow-based estimates are correct and accurate. Our work therefore clearly demonstrates that flow models can approximate single states of interest with high accuracy without training data, providing a solid foundation for follow-up work.

\update{A current limitation of our proposed method is the computational cost of training. Although generating samples from the model and obtaining their probability density is efficient as it is trivially parallelizable, training the model with gradient-based methods is inherently sequential. While training took only a day on the smallest system ($64$-particle mW), reaching convergence of the free energy estimates for the biggest systems ($512$-particle mW and $500$-particle LJ\@)  took approximately 3 weeks on 16 A100 GPUs (details in the Supplementary Material). Therefore, our approach is best suited for applications where the cost of training can be amortized across several evaluations. In particular, a promising research direction is training a single model parameterized by state variables or order parameters (such as temperature, pressure, particle density, etc.), so that a range of states or systems can be approximated at the cost of training only once.}

With rapid improvements in model architectures, optimizers and training schemes, it seems likely that this type of approach can be scaled up to larger system sizes and other types of challenging systems, such as explicit water models with electrostatic interactions and rotational degrees of freedom, in the future. The application and adaptation of increasingly suitable normalizing flows is a very active area of research, for example, the concurrent work on solids in Ref.~\cite{Rasool2021}, yielding increasingly flexible flows for progressively more general systems. The end-to-end differentiability of our approach (or one more general) could be leveraged to address difficult inverse material design problems, extending recent MD-based approaches~\cite{Goodrich2021}.

\ack
\ifanonymized
Removed for double-anonymous review.
\else
We would like to thank our colleagues Stuart Abercrombie, Danilo Jimenez Rezende, Th{\'e}ophane Weber, Daan Wierstra, Arnaud Doucet, Peter Battaglia, James Kirkpatrick, John Jumper, Alex Goldin and Guy Scully for their help and for stimulating discussions.
\fi

\section*{References}
\bibliographystyle{iopart-num}
\bibliography{references}

\providecommand{\newblock}{}
\begin{thebibliography}{10}
\expandafter\ifx\csname url\endcsname\relax
  \def\url#1{{\tt #1}}\fi
\expandafter\ifx\csname urlprefix\endcsname\relax\def\urlprefix{URL }\fi
\providecommand{\eprint}[2][]{\url{#2}}

\bibitem{Tuckerman2019}
Tuckerman M~E 2019 {\em Science\/} {\bf 365} 982--983

\bibitem{FrenkelSmit}
Frenkel D and Smit B 2002 {\em Understanding Molecular Simulation\/} 2nd ed
  (San Diego: Academic Press) ISBN 978-0-12-267351-1

\bibitem{Duane1987}
Duane S, Kennedy A, Pendleton B~J and Roweth D 1987 {\em Phys. Lett. B\/} {\bf
  195} 216--222

\bibitem{Yu2016}
Yu I, Takaharu M, Ando T, Harada R, Jung J, Sugita Y and Feig M 2016 {\em eLife
  2016;5:e19274\/}

\bibitem{Hudait2017}
Lupi L, Hudait A, Peters B, Gr\"{u}nwald M, Gotchy~Mullen R, Nguyen A~H and
  Molinero V 2017 {\em Nature\/} {\bf 551} 218--222

\bibitem{Mosalaganti2021}
Mosalaganti S, Obarska-Kosinska A, Siggel M, Turonova B, Zimmerli C~E, Buczak
  K, Schmidt F~H, Margiotta E, Mackmull M~T, Hagen W, Hummer G, Beck M and
  Kosinski J 2021 {\em bioRxiv:2021.10.26.465776\/}

\bibitem{Tabak2013}
Tabak E~G and Turner C~V 2013 {\em Commun. Pure Appl. Math.\/} {\bf 66}
  145--164

\bibitem{Rezende2015}
Rezende D~J and Mohamed S 2015 Variational inference with normalizing flows
  {\em 32nd Int. Conf. Mach. Learn.\/} pp 1530--1538

\bibitem{Albergo2019}
Albergo M~S, Kanwar G and Shanahan P~E 2019 {\em Phys. Rev. D\/} {\bf 100}(3)
  034515

\bibitem{Boyda2021}
Boyda D, Kanwar G, Racani\`ere S, Rezende D~J, Albergo M~S, Cranmer K, Hackett
  D~C and Shanahan P~E 2021 {\em Phys. Rev. D\/} {\bf 103}(7) 074504

\bibitem{Nicoli2021}
Nicoli K~A, Anders C~J, Funcke L, Hartung T, Jansen K, Kessel P, Nakajima S and
  Stornati P 2021 {\em Phys. Rev. Lett.\/} {\bf 126}(3) 032001

\bibitem{Nicoli2020}
Nicoli K~A, Nakajima S, Strodthoff N, Samek W, M\"uller K~R and Kessel P 2020
  {\em Phys. Rev. E\/} {\bf 101}(2) 023304

\bibitem{Noe2019}
No{\'e} F, Olsson S, K{\"o}hler J and Wu H 2019 {\em Science\/} {\bf 365}
  eaaw1147

\bibitem{Papamakarios2021}
Papamakarios G, Nalisnick E, Rezende D~J, Mohamed S and Lakshminarayanan B 2021
  {\em J. Mach. Learn. Res.\/} {\bf 22} 1--64

\bibitem{Kobyzev2021}
Kobyzev I, Prince S~J and Brubaker M~A 2021 {\em IEEE Trans. Pattern Anal.
  Mach. Intell.\/} {\bf 43} 3964--3979

\bibitem{Bugallo2017}
Bugallo M~F, Elvira V, Martino L, Luengo D, Miguez J and Djuric P~M 2017 {\em
  IEEE Signal Process. Mag.\/} {\bf 34} 60--79

\bibitem{Muller2019}
M\"{u}ller T, McWilliams B, Rousselle F, Gross M and Nov\'{a}k J 2019 {\em ACM
  Trans. Graph.\/} {\bf 38}

\bibitem{Shirts2008}
Shirts M~R and Chodera J~D 2008 {\em J. Chem. Phys.\/} {\bf 129} 124105

\bibitem{Jarzynski2002}
Jarzynski C 2002 {\em Phys. Rev. E\/} {\bf 65}(4) 046122

\bibitem{Hahn2009}
Hahn A~M and Then H 2009 {\em Phys. Rev. E\/} {\bf 79} 011113

\bibitem{Wirnsberger2020}
Wirnsberger P, Ballard A~J, Papamakarios G, Abercrombie S, Racani{\`e}re S,
  Pritzel A, Rezende D~J and Blundell C 2020 {\em J. Chem. Phys.\/} {\bf 153}
  144112

\bibitem{Ding2020}
Ding X and Zhang B 2020 {\em J. Phys. Chem. B\/} {\bf 124} 10166--10172

\bibitem{Rizzi2021}
Rizzi A, Carloni P and Parrinello M 2021 {\em J. Phys. Chem. Lett.\/} {\bf 12}
  9449--9454

\bibitem{Xinqiang2021}
Ding X and Zhang B 2021 {\em J. Phys. Chem. Lett.\/} {\bf 12} 2509--2515

\bibitem{Kohler2020}
K{\"o}hler J, Klein L and No{\'e} F 2020 Equivariant flows: Exact likelihood
  generative learning for symmetric densities {\em 37th Int. Conf. Mach.
  Learn.\/} pp 5361--5370

\bibitem{Molinero2009}
Molinero V and Moore E~B 2009 {\em J. Phys. Chem. B\/} {\bf 113} 4008--4016

\bibitem{Rezende2020}
Rezende D~J, Papamakarios G, Racani{\`{e}}re S, Albergo M~S, Kanwar G, Shanahan
  P~E and Cranmer K 2020 Normalizing flows on tori and spheres {\em 37th Int.
  Conf. Mach. Learn.\/} pp 8083--8092

\bibitem{Vaswani2017}
Vaswani A, Shazeer N, Parmar N, Uszkoreit J, Jones L, Gomez A~N, Kaiser L and
  Polosukhin I 2017 Attention is all you need {\em Adv. Neural Inf. Process.
  Syst.\/}

\bibitem{Bender2020}
Bender C~M, O'Connor K, Li Y, Garcia J~J, Oliva J~B and Zaheer M 2020
  Exchangeable generative models with flow scans {\em 34th AAAI Conf. Artif.
  Intell.\/}

\bibitem{Aragones2012}
Aragones J~L, Valeriani C and Vega C 2012 {\em J. Chem. Phys.\/} {\bf 137}
  146101

\bibitem{Quigley2014}
Quigley D 2014 {\em J. Chem. Phys.\/} {\bf 141} 121101

\bibitem{jax2018github}
Bradbury J, Frostig R, Hawkins P, Johnson M~J, Leary C, Maclaurin D, Necula G,
  Paszke A, Vander{P}las J, Wanderman-{M}ilne S and Zhang Q 2018 {JAX}:
  composable transformations of {P}ython+{N}um{P}y programs
  \urlprefix\url{http://github.com/google/jax}

\bibitem{haiku2020github}
Hennigan T, Cai T, Norman T and Babuschkin I 2020 {H}aiku: {S}onnet for {JAX}
  \urlprefix\url{http://github.com/deepmind/dm-haiku}

\bibitem{deepmind2020jax}
Babuschkin I, Baumli K, Bell A, Bhupatiraju S, Bruce J, Buchlovsky P, Budden D,
  Cai T, Clark A, Danihelka I, Fantacci C, Godwin J, Jones C, Hennigan T,
  Hessel M, Kapturowski S, Keck T, Kemaev I, King M, Martens L, Mikulik V,
  Norman T, Quan J, Papamakarios G, Ring R, Ruiz F, Sanchez A, Schneider R,
  Sezener E, Spencer S, Srinivasan S, Stokowiec W and Viola F 2020 The
  {D}eep{M}ind {JAX} {E}cosystem \urlprefix\url{http://github.com/deepmind}

\bibitem{Thompson2022}
Thompson A~P, Aktulga H~M, Berger R, Bolintineanu D~S, Brown W~M, Crozier P~S,
  {in 't Veld} P~J, Kohlmeyer A, Moore S~G, Nguyen T~D, Shan R, Stevens M~J,
  Tranchida J, Trott C and Plimpton S~J 2022 {\em Comput. Phys. Commun.\/} {\bf
  271} 108171

\bibitem{Lechner2008}
Lechner W and Dellago C 2008 {\em J. Chem. Phys.\/} {\bf 129} 114707

\bibitem{Auer2001}
Auer S and Frenkel D 2001 {\em Nature\/} {\bf 409} 1020--1023

\bibitem{Vega2007}
Vega C and Noya E~G 2007 {\em J. Chem. Phys.\/} {\bf 127} 154113

\bibitem{Bennett1976}
Bennett C~H 1976 {\em J. Comp. Phys.\/} {\bf 22} 245--268

\bibitem{Shirts2003}
Shirts M~R, Bair E, Hooker G and Pande V~S 2003 {\em Phys. Rev. Lett.\/} {\bf
  91}(14) 140601

\bibitem{Frenkel1984}
Frenkel D and Ladd A~J~C 1984 {\em J. Chem. Phys.\/} {\bf 81} 3188--3193

\bibitem{Bruce1997}
Bruce A~D, Wilding N~B and Ackland G~J 1997 {\em Phys. Rev. Lett.\/} {\bf
  79}(16) 3002--3005

\bibitem{Rasool2021}
Ahmad R and Cai W 2021 {\em arXiv:2111.01292\/}

\bibitem{Goodrich2021}
Goodrich C~P, King E~M, Schoenholz S~S, Cubuk E~D and Brenner M~P 2021 {\em
  Proc. Natl. Acad. Sci. U.S.A.\/} {\bf 118} e2024083118

\bibitem{Kingma2014}
Kingma D~P and Ba J 2015 Adam: A method for stochastic optimization {\em 3rd
  Int. Conf. Learn. Represent.\/}

\bibitem{Polson2000}
Polson J~M, Trizac E, Pronk S and Frenkel D 2000 {\em J. Chem. Phys.\/} {\bf
  112} 5339--5342

\bibitem{Vega2008}
Vega C, Sanz E, Abascal J~L~F and Noya E~G 2008 {\em J. Phys.: Condens.
  Matter\/} {\bf 20} 153101

\bibitem{Beutler1994}
Beutler T~C, Mark A~E, {van Schaik} R~C, Gerber P~R and {van Gunsteren} W~F
  1994 {\em Chem. Phys. Lett.\/} {\bf 222} 529--539

\end{thebibliography}

\ifincludesupplement
\newpage
{\noindent\Large{\textbf{Supplementary Material}}}
\newcommand{\beginsupplement}{%
        \setcounter{table}{0}
        \renewcommand{\tablename}{Supplementary Table}
        \setcounter{figure}{0}
        \renewcommand{\theequation}{S\arabic{equation}}
        \setcounter{equation}{0}
        \renewcommand{\figurename}{Supplementary Figure}
        \setcounter{section}{0}
        \renewcommand\thesection{\Alph{section}}
     }
\beginsupplement

\section{Details of flow architecture}

\begin{figure}[h]
\centering
\includegraphics[width=0.9\textwidth]{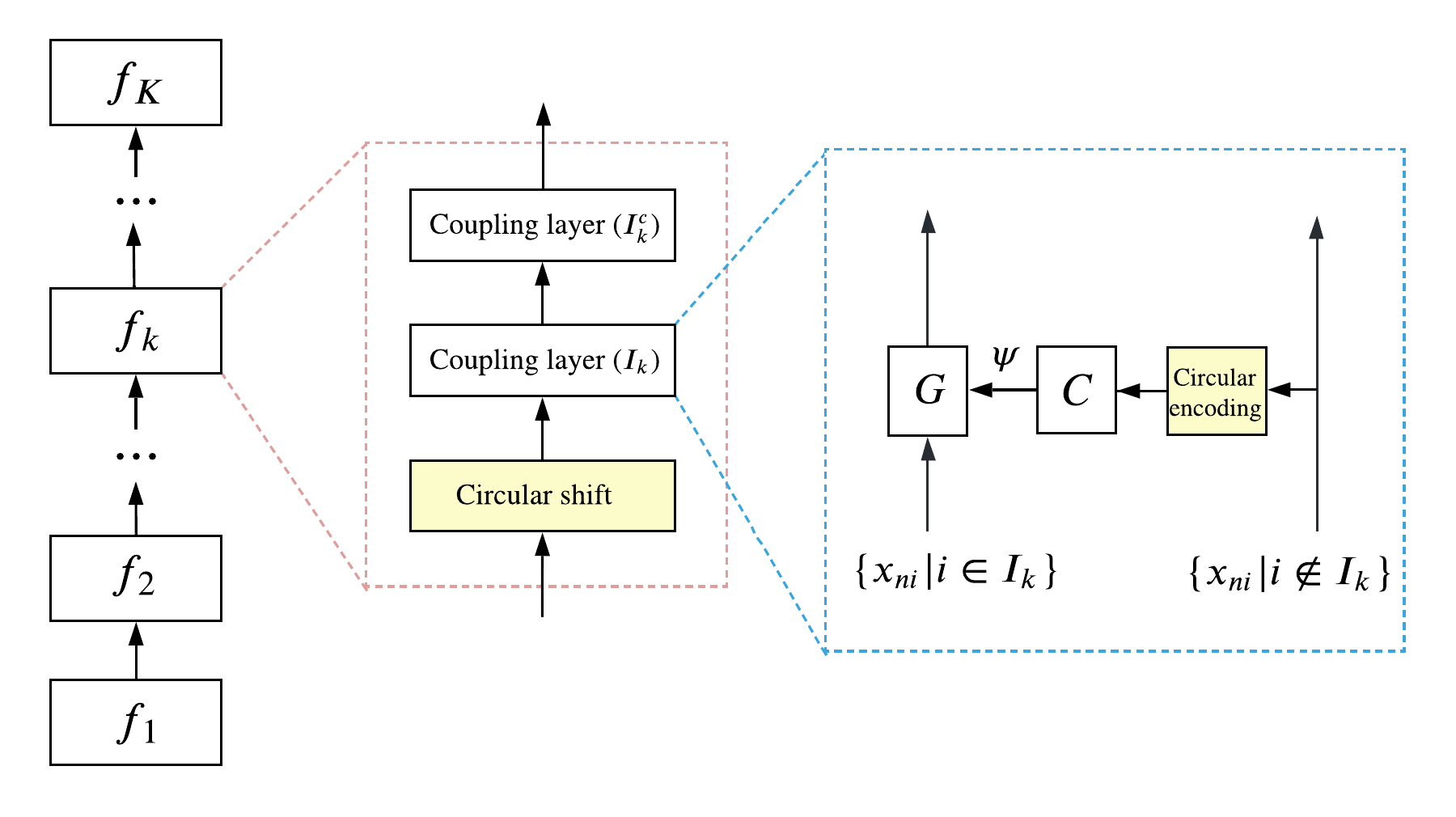}
\caption{\label{smfig:flow_architecture}Flow architecture, based on Ref.~\cite{Wirnsberger2020}. Improvements are highlighted in yellow (learned circular shift and higher-order circular encoding).}
\end{figure}

Our normalizing flow $f = f_K\circ\cdots\circ f_1$ is an improved version of the implementation proposed in Ref.~\cite{Wirnsberger2020}, which is a sequence of permutation-equivariant coupling layers and is illustrated in Fig.~\ref{smfig:flow_architecture}.
Given a subset $I_k\subset\set{1, 2, 3}$ of particle coordinates, a coupling layer is an invertible mapping that transforms a collection of $N$ particles as follows:
\begin{equation}
    x_{ni}\mapsto\begin{cases}
    G(x_{ni}; \psi_{ni}) & i \in I_k \\
    x_{ni} & i \notin I_k.
    \end{cases}
\end{equation}
That is, particle coordinates indexed by $I_k$ are transformed element-wise by a transformation $G$ parameterized by $\psi$, whereas the remaining coordinates stay fixed. The parameters $\psi$ are computed as a function $C$ of all the non-transformed coordinates, that is, $\psi = C(\set{x_{ni}\,|\,i\notin I})$, which introduces dependencies between particles. Each flow layer $f_k$ is the composition of two coupling layers with complementary index sets $I_k$ and $I_k^c = \set{1, 2, 3}\setminus I_k$, so that the particle coordinates which stay fixed in one coupling layer are transformed by the other. In addition, different subsets $I_k$ are used in different flow layers $f_k$, so that all possible coordinate splits are cycled over. The element-wise transformation $G$ is a circular rational-quadratic spline~\cite{Rezende2020}, which is a non-linear invertible transformation that respects boundary conditions. The function $C$ is a transformer~\cite{Vaswani2017} in a permutation-equivariant configuration, which ensures that the flow $f$ is equivariant to particle permutations.

We refer the reader to Section V of Ref.~\cite{Wirnsberger2020} for further details on the flow architecture. The specific improvements we make in this paper are the following.
\begin{itemize}
    \item The coupling layers used in Ref.~\cite{Wirnsberger2020} have the property that particle coordinates at the edge of the box remain fixed, which limits the flexibility of the flow. To overcome this limitation, we interleave the flow layers $f_k$ with learned circular shifts, defined by
    \begin{equation}
        x_{ni} \mapsto (x_{ni} + c_i) \text{ mod } (L_i/\sigma),
    \end{equation}
    were $c_i$ is a learned parameter corresponding to dimension $i=1, 2, 3$. The parameter $c_i$ is a constant---it does not depend on any of the coordinates---so the Jacobian determinant of the above transformation is equal to $1$. 
    \item Reference~\cite{Wirnsberger2020} uses a circular encoding of the coordinates prior to feeding them to the transformer in order to encode their periodicity due to periodic boundary conditions. Specifically, a coordinate $x_{ni}$ is encoded as
    \begin{equation}
    x_{ni} \mapsto [\cos\brk{\omega_i x_{ni}}, \sin\brk{\omega_i x_{ni}}],
\end{equation}
    where $\omega_i = \frac{2\pi }{L_i/\sigma}$.
    Here we use a richer encoding that also includes higher-order frequencies, and is defined by
    \begin{align}
    x_{ni} \mapsto [&\cos\brk{\omega_i x_{ni}},
    \sin\brk{\omega_i x_{ni}},\\
    &\cos\brk{2\omega_i x_{ni}}, \sin\brk{2\omega_i x_{ni}},\\
    &\ldots,\\
    &\cos\brk{N_{f}\omega_i x_{ni}}, \sin\brk{N_f\omega_i x_{ni}}],
\end{align}
where $N_f$ is the total number of frequencies, a hyperparameter to be tuned. Although the higher frequency inputs could be deduced from the lowest frequency, we found that the higher frequencies significantly helped the network express more flexible functions.
\end{itemize}
Table~\ref{tab:model_hyper} lists the hyperparameters we used for the experiments in this paper.

\begin{table}[ht]
\caption{\label{tab:model_hyper}Model hyperparameters.}
\begin{tabularx}{\textwidth}{XX} 
\toprule
\textbf{Normalizing flow} & \\ 
Number of layers ($K$) & $24$ \\
\\
\textbf{Transformer} & \\ 
Number of blocks & $2$ \\
Number of heads &  $2$ \\
Embedding dimension & $256$ \\
Number of frequencies in circular encoding ($N_f$) & 
$8$ (mW $64$ particles), $16$ (mW $216$ and LJ $256$ particles), 24 (mW $512$ particles), 32 (LJ $500$ particles) \\
\\
\textbf{Circular rational-quadratic spline} & \\ 
Number of segments & $16$\\
\\
\textbf{Base distribution} & \\ 
Standard deviation of truncated Gaussian noise (in reduced units)
&
$0.01$ (LJ), $0.08$ (mW)
\\
\bottomrule
\end{tabularx}
\end{table}

\section{Optimization details}

All models were trained using the Adam optimizer~\cite{Kingma2014}. Prior to applying the Adam update rule, we clip the norm of the gradient to be below a maximum value, to avoid potential instability due to large gradients. Also, we reduce the learning rate during training at pre-specified training steps by a constant factor. Table~\ref{tab:train_hyper} lists the training hyperparameters we used in our experiments.

The pairwise potentials we use in the experiments (the Lennard-Jones potential and the two-body term of the monatomic Water potential) have the property that they diverge for zero pairwise distance. This can cause numerical problems during training, as it can cause the gradients of the loss function to become very large. To avoid this problem, we train on a linearized version of these potentials, defined by:
\begin{equation}
    u_{\mathrm{lin}}(r) = \begin{cases} 
    u(r_{\mathrm{lin}}) + u'(r_{\mathrm{lin}})\brk{r - r_{\mathrm{lin}}} & r < r_{\mathrm{lin}} \\
u(r) & r\ge r_{\mathrm{lin}}, \end{cases}
\end{equation}
where $u$ is the original pairwise potential and $r_{\mathrm{lin}}$ is a distance threshold below which it is linearized. In practice, we set $r_{\mathrm{lin}}$ smaller than the typical distance between particles so linearization has a negligible effect on the target Boltzmann distribution.

\begin{table}[ht]
\caption{\label{tab:train_hyper}Training hyperparameters.}
\begin{tabularx}{\textwidth}{XX}
\toprule
\textbf{General}\\
Batch size & $128$ \\
Maximum gradient norm & $10^4$\\
\\
\textbf{Learning-rate schedule}\\
Initial learning rate & $7\cdot10^{-5}$\\
Learning-rate decay steps & $250$k, $500$k\\
Learning-rate decay factor & $0.1$\\
\\
\textbf{Energy linearization}\\
Distance threshold to linearize below ($r_{\mathrm{lin}}/\sigma$)  & $0.8$ (LJ), $0.5$ (mW)\\ 
\\
\textbf{Adam optimizer}\\
$\beta_1$  & $0.9$\\ 
$\beta_2$  & $0.999$\\
$\epsilon$  & $10^{-8}$\\
\bottomrule
\end{tabularx}
\end{table}

\section{Physical systems under study}

We apply our method to two particular physical systems in the solid state: truncated and shifted Lennard-Jones (LJ) and monatomic Water (mW). We now summarize the energy functions of these systems.

\subsection{Lennard-Jones}
The pairwise LJ potential is given by~\cite{FrenkelSmit}
\begin{equation}
    u_\text{LJ}(r) = 4\epsilon \left[ {\left(  \frac{\sigma}{r} \right)}^{12}  - {\left(  \frac{\sigma}{r} \right)}^{6} \right],
\end{equation}
where $r$ is the distance between two particles, $\epsilon$ defines the unit of energy and the particle diameter $\sigma$ the unit of length. To compare free energy estimates with literature values reported in Ref.~\cite{Aragones2012}, we employed a spherically truncated and shifted version of the above potential given by~\cite{FrenkelSmit}
\begin{equation}
    u_\text{tr-sh}(r) = 
    \begin{cases}
        u_\text{LJ}(r)   - u_\text{LJ}(r_\text{c})          & \text{if   }  r \leq r_\text{c}, \\
        0                                                   & \text{otherwise,}
\end{cases}
\end{equation}
with $r_\text{c}$ being a radial cutoff set to $2.7\sigma$. The total potential energy comprises contributions over all pairs of the $N$-particle system described by $R=(r_1, \ldots, r_N)$, where $r_n = (r_{n1}, r_{n2}, r_{n3})$ defines the position of the $n$-th atom with coordinates $r_{n\alpha} \in [0, L_\alpha]$, and is given by
\begin{equation}
    U_\text{LJ}(R) = \sum_i \sum_{j > i} u_\text{tr-sh}(|{\tilde r}_{ij}|),
\end{equation}
where ${\tilde r}_{ij} = \text{pbc}(r_j - r_i)$ and the function $\text{pbc}$ is applied elementwise yielding the shortest, signed distance for the component $y_\alpha$ of its argument, that is $\text{pbc}(y_\alpha) = y_\alpha - L_{\alpha} \text{round}(y_\alpha /L_{\alpha})$. Finally, we switch to reduced units by choosing $\sigma$ as the unit of length and $\epsilon$ as the unit of energy, and establish the connection with $x = R/\sigma$ and $U(x) = U(R)/\epsilon$ that we introduced in the main text. See Ref.~\cite{FrenkelSmit} for further details on reduced units.

\subsection{Monatomic Water}
Monatomic Water~\cite{Molinero2009} models water as point particles interacting via two-body and three-body potentials. The total energy is given by
\begin{equation}
    U_\text{mW}(R) =\sum_i \sum_{j > i}  \phi_1({\tilde r}_{ij}) +  \sum_i \sum_{j \neq  i} \sum_{k > j} \phi_2({\tilde r}_{ij}, {\tilde r}_{ik}, \theta_{ijk})
\end{equation}
where $\theta_{ijk}$ is an angle computed for each triplet, and the two functions are
\begin{equation}
  \phi_1(r) = A \epsilon \left[ B   {\left(  \frac{\sigma}{r} \right)}^{4} - 1 \right] \exp\left(\frac{\sigma}{r - a \sigma}\right)
\end{equation}
and
\begin{equation}
  \phi_2(r,s, \theta) = \lambda \epsilon {\left( \cos\theta - \cos\theta_0  \right)}^2 \exp{ \left(\frac{\gamma \sigma}{r - a \sigma}\right)} \exp{\left(\frac{\gamma \sigma}{s - a \sigma}\right)}
\end{equation}
with $A = 7.049556277$, $B = 0.6022245584$, $a = 1.8$, $\lambda = 23.15$, $\theta = 109.47^\circ$ and $\gamma = 1.2$. The remaining two parameters $\epsilon = 6.189~\text{kcal/mol}$ and $\sigma = 2.3925~\text{\AA}$ set the scales for energy and length, similarly to the case of LJ above.

\section{MD simulation details}

We used molecular dynamics (MD) to simulate the above systems in the canonical $NVT$ ensemble: $N$ particles in a fully periodic simulation box of volume $V=L_1 L_2 L_3$ and temperature $T$. 

All MD simulations were carried out with the package LAMMPS~\cite{Thompson2022}, defined by the following parameters (see Table~\ref{tab:sim} for values). To simulate at constant temperature, we used a Langevin thermostat with damping constant $\tau$ and subtracted a force for the centre of mass to remain stationary (keyword \enquote{zero yes}). The equations of motion were integrated using the velocity Verlet algorithm and discretized using a timestep $\Delta t$. We initialized particle positions with the target lattice and performed an equilibration run of length $t_\text{run}$ followed by a production run of the same length during which we sampled particle positions every $M$ timesteps.

\begin{table}
\caption{\label{tab:sim}
Simulation parameters. For LJ, all quantities are reported in reduced units.}
\begin{tabularx}{\textwidth}{XXXp{3cm}XXXXp{3cm}X}
\toprule
    &  $\Delta t$ &   $T$    &  $\rho$                        &  $\tau$  &  $t_\text{run}$  &  $M$     & $N_\lambda$ & $\Lambda_E$ \\ \midrule
LJ  &  $10^{-4}$  &   2      &  1.28                          &  0.2     &  $10^4$          & $10^4$  & 100  & 2500  \\
mW  &  0.1~fs     &   200~K  &  $0.033567184~\text{\AA}^{-3}$ &  1~ps    &  10~ns           & $10^4$  & 200  &10 $\text{kcal} / (\text{mol}\ \text{\AA}^{2})$\\  
\bottomrule
\end{tabularx}
\end{table}

\section{MBAR details}
For comparison with the flow-based free energy estimates, we compute the value of the solid free energy using the multistate Bennett acceptance ratio (MBAR) method~\cite{Shirts2008}. To this end, we decomposed the free energy $F$ into two contributions, $F = F_0 + \Delta F$. The first term is the free energy of an Einstein crystal, which is approximated as~\cite{Polson2000, Vega2008}
\begin{equation}
\frac{\beta F_\text{0}}{N} = \frac{1}{N} \ln\left(\frac{N \Lambda^3}{V}\right) 
+ \frac 32 \left(1 - \frac{1}{N}\right) \ln \left(\frac{\beta \Lambda_E \Lambda^2}{\pi}\right)
- \frac{3}{2N} \ln N,
\end{equation}
where $\Lambda$ is the thermal de Broglie wavelength, which we set to the particle diameter $\sigma$ following Ref.~\cite{Vega2007}, $\beta = 1/k_\text{B} T$ is the inverse temperature and $k_\text{B}$ is the Boltzmann constant. The quantity $\Lambda_E$ defines the spring constant of the Einstein crystal with energy
\begin{equation}
    U_\text{id}(R)  = \Lambda_E \sum_{i = 1}^N {\left|r_i - r_{i}^0\right|}^2,
\end{equation}
where $r_{i}^0$ denotes the position of the lattice site with which particle $i$ is associated. To estimate $\Delta F$, we introduced intermediate energies $U(R; \lambda)$ for each system (see below) to interpolate between the ideal Einstein crystal defined by $U_\text{id}(R)$ and the target solid based on a scalar parameters $\lambda \in [0, 1]$. We then discretized $\lambda$ uniformly into $N_\lambda$ values and performed intermediate simulations for each value of $\lambda$ during which we collected 10k samples per simulation. The energy matrix computed from the samples was then used as input to MBAR~\cite{Shirts2008}. We repeated this procedure to obtain estimates for ten different random seeds, and reported the mean and two standard deviation across the seed estimates in Tab.~\ref{tab:fe} of the main text.

\subsection{Interpolation stages for Lennard-Jones system}
We use a softcore lambda version of the LJ potential $U_\text{soft}(R; \lambda)$ to interpolate between $U_\text{soft}(R; \lambda=0) = 0$ and $U_\text{soft}(R; \lambda = 1) = U_\text{LJ}(R)$~\cite{Beutler1994}. We then constructed the energy $U(R; \lambda)  = U_\text{soft}(R; \lambda) + (1-\lambda)  U_\text{id}(R)$ for the simulations. 

\subsection{Interpolation stages for monatomic Water}
For ice, we employed a simple linear interpolation $U(R; \lambda)  = \lambda U_\text{mW}(R) + (1-\lambda)  U_\text{id}(R).$

\section{Hardware details and computational cost}

For our flow experiments, we used 16 A100 GPUs to train each model on the bigger systems ($512$-particle mW and $500$-particle LJ\@). It took approximately 3 weeks of training to reach convergence of the free-energy estimates. Obtaining 2M samples for evaluation took approximately 12 hours on 8 V100 GPUs for each of these models. Training on the medium-size systems ($216$-particle mW and $256$-particle LJ) until convergence took about 4 days on 8 V100 GPUs, obtaining 2M samples took about 4 hours on 4 V100 GPUs. The $64$-particle mW systems trained in one day on 4 V100 GPUs and 2M samples took 2 hours to generate on the same 4 V100 GPUs.

For each baseline MBAR estimate, we performed 100 separate simulations for LJ and 200 for mW, corresponding to the number of stages employed. These simulations were performed with LAMMPS~\cite{Thompson2022} and each of them ran on multiple CPU cores communicating via MPI\@. We used 4 cores for the $64$-particle and $216$-particle mW experiments and 8 cores for all other systems. The MD simulations completed after approximately 11 and 14 hours for LJ (256 and 500 particles), and 7, 20 and 48 hours for mW (64, 216 and 512 particles). To evaluate the energy matrix for a single MBAR estimate, we decomposed the problem into the number of stages separate jobs (100 for LJ and 200 for mW), so that each worker evaluated all energies for the samples corresponding to a single stage on a V100 GPU\@. Each of these jobs took less than 10 minutes for LJ (both system sizes) and approximately 0.5, 1 and 4 hours for mW with 64, 216 and 512 particles. Running pymbar~\cite{Shirts2008} until convergence on a CPU took between 20 minutes and two hours for a single estimate.

\section{Supplementary experimental results}

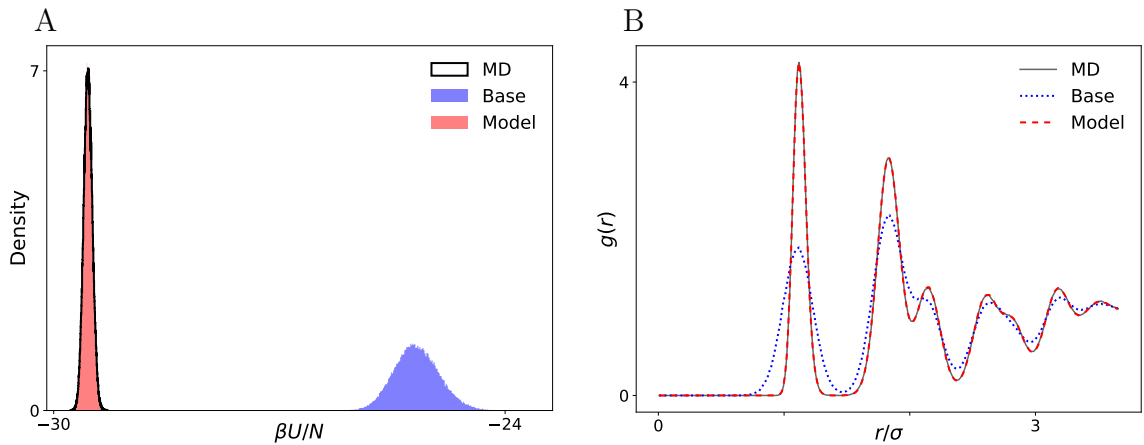
\begin{figure}[ht]
\centering
    \begin{tikzpicture} [every node/.style={inner sep=0,outer sep=-1}]
        \draw (0, 0) node[inner sep=0] {\includegraphics[width=.49\textwidth]{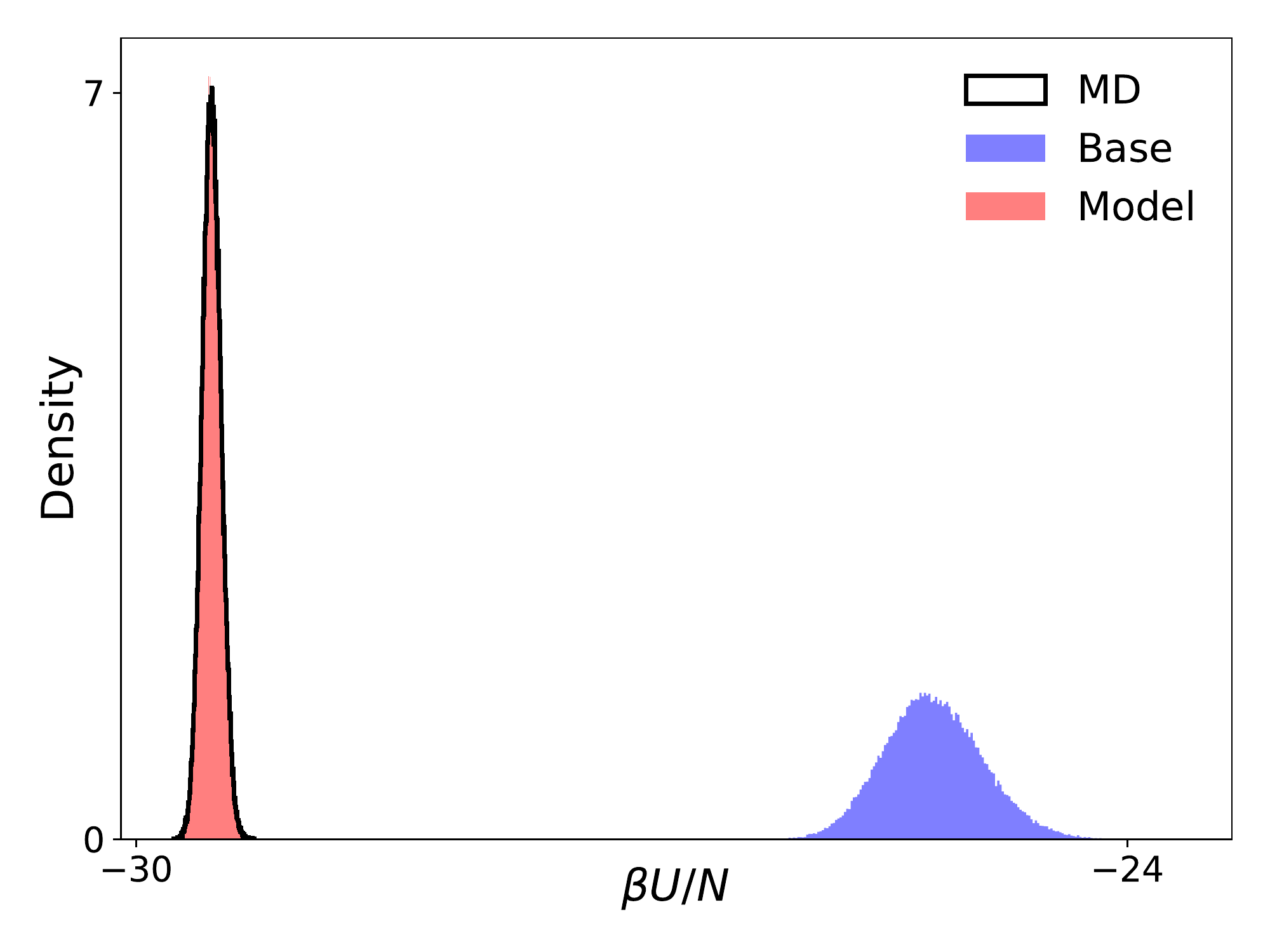}};
        \coordinate (A) at (current bounding box.north west);
        \node[above right, xshift=0.6cm] at (A) {A};
    \end{tikzpicture}
    \begin{tikzpicture} [every node/.style={inner sep=0,outer sep=-1}]
        \draw (0, 0) node[inner sep=0] {\includegraphics[width=.49\textwidth]{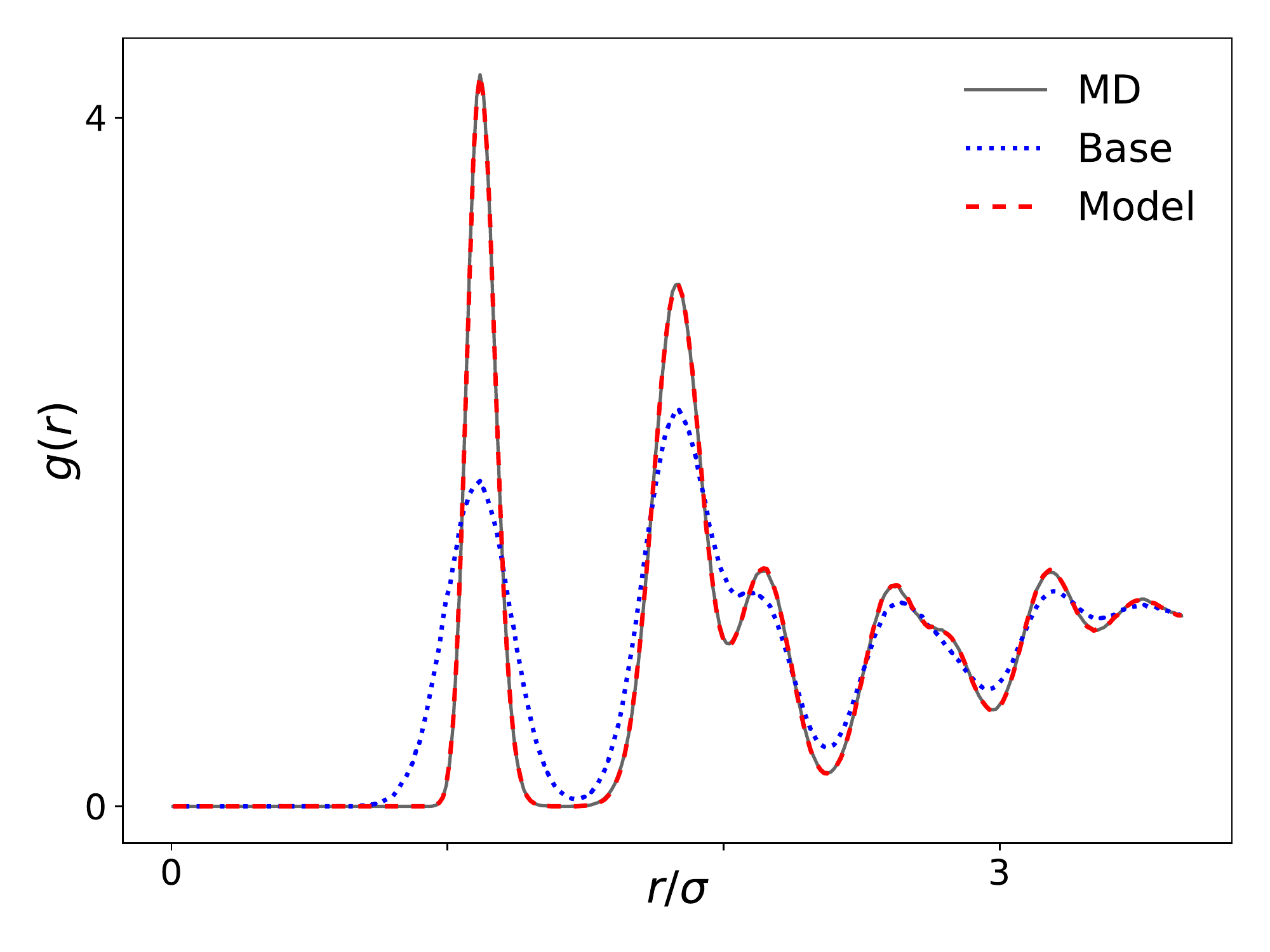}};
        \coordinate (B) at (current bounding box.north west);
        \node[above right, xshift=0.6cm] at (B) {B};
    \end{tikzpicture}
\caption{\label{smfig:rdf_hex}Energy histograms (A) and radial distribution functions (B) of the base distribution, the fully trained model and MD simulation data, for the 512-particle hexagonal ice system.}
\end{figure}

\begin{figure}
\centering
\includegraphics[width=0.6\textwidth]{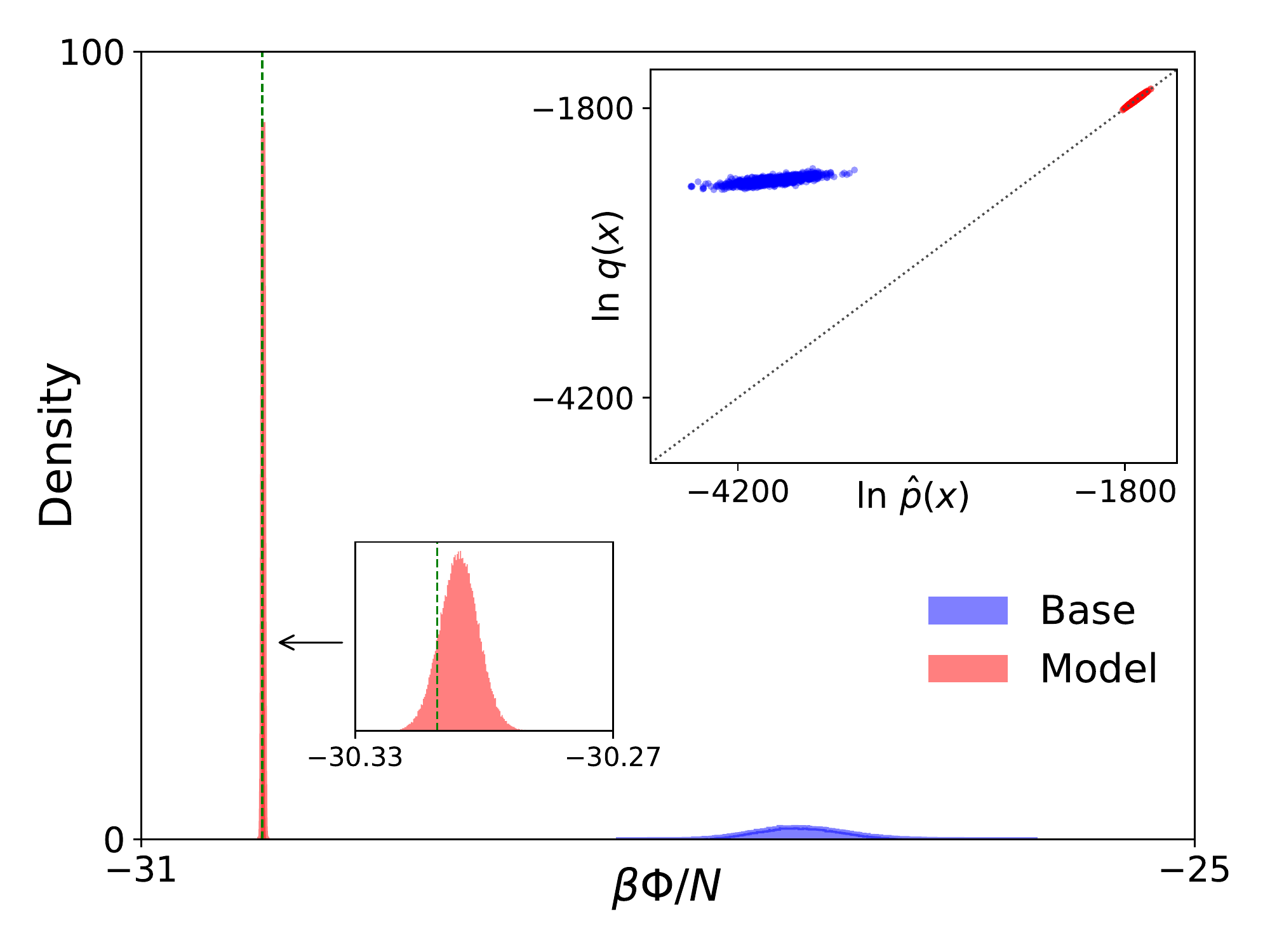}
\caption{\label{smfig:energy_hex} Histograms of work values ($\beta\Phi$) per particle from base and model samples for 512-particle hexagonal ice; the vertical line marks $-\widehat{\nlog{Z}}/N$ as estimated by MBAR (enlarged in inset). Top right inset: scatter plot of model density vs approximately normalized target density $\ln \hat{p} = -\beta U - \widehat{\nlog{Z}}$ computed from model samples, where the model is either the base distribution or the fully trained model and $\widehat{\nlog{Z}}$ is the MBAR estimate. \update{The dotted diagonal marks the identity.}}
\end{figure}

\else
\fi

\end{document}